\documentclass[twocolumn,pre,showpacs,amsmath,longbibliography]{revtex4-1}
\usepackage[dvipdfmx]{graphicx}
\usepackage{url}
\usepackage{cases}
\usepackage{bm}
\usepackage[dvipsnames]{xcolor}
\usepackage[version=3]{mhchem}
\usepackage[normalem]{ulem}
\usepackage[hidelinks]{hyperref}
\usepackage{here}
\usepackage{booktabs}
\usepackage{multirow}

\newcommand{\bra}{\left\langle}
\newcommand{\ket}{\right\rangle}

\newcommand{\bv}[1]{{\boldsymbol #1}}

\newcommand{\pder}[2]{\frac{\partial #1}{\partial  #2}}

\newcommand{\abs}[1]{\left|#1\right|}
\newcommand{\sbkt}[1]{\langle#1\rangle}

\newcommand{\kB}{k_{\rm B}}

\newcommand{\vep}{\varepsilon}
\newcommand{\eq}{\mathrm{eq}}

\newcommand{\subL}{\mathrm{L}}
\newcommand{\subG}{\mathrm{G}}

\newcommand{\mT}{T_\mathrm{m}}

\newcommand{\HT}{T_\mathrm{H}}
\newcommand{\LT}{T_\mathrm{C}}
\newcommand{\Tm}{T_\mathrm{m}}

\newcommand{\kinT}{T(x)}

\begin{document}

\title{Heat-induced liquid hovering in liquid-gas coexistence under gravity}

\author{Akira Yoshida}
\email{a.yoshida.phys@gmail.com}
\affiliation {Department of Physics,  Ibaraki University, Mito 310-8512, Japan}

\author{Naoko Nakagawa}
\email{naoko.nakagawa.phys@vc.ibaraki.ac.jp}
\affiliation {Department of Physics,  Ibaraki University, Mito 310-8512, Japan}

\author{Shin-ichi Sasa}
\email{sasa.shinichi.6n@kyoto-u.ac.jp}
\affiliation {Department of Physics,  Kyoto University, Kyoto 606-8502, Japan}
\date{\today}

\begin{abstract}
  We study a liquid-gas coexistence system in a container under gravity with heat flow in the direction opposite to gravity. By molecular dynamics simulation, we find that the liquid buoys up and continues to float steadily. The height at which the liquid floats is determined by a dimensionless parameter related to the ratio of the temperature gradient to gravity. We confirm that supercooled gas remains stable above the liquid. We provide a phenomenological  argument for explaining the phenomenon from a simple thermodynamic assumption.
\end{abstract}

\maketitle

{\em Introduction.---} 
Clouds float in the sky although they are heavier than the surface air. According to the standard theory, thermal convection caused by the temperature difference between the sea surface and the atmosphere forms cumulonimbus clouds hovering over the sea \cite{agee1987,orville1996,dror2023}. While cloud formation may involve various complex processes, convection certainly plays a significant role in this phenomenon. The Leidenfrost effect is another phenomenon of a liquid floating above a gas \cite{leidenfrost1756,quere2013,quere2019,rodrigues2019}. When we drop a droplet onto a hot plate, heat transfers violently from the plate to the droplet. The droplet then evaporates instantly, and the vapor envelops the droplets and causes them to float. Even in this case, a complex flow of materials appears. 

Motivated by these phenomena, we investigate the possibility that a liquid can float over a gas against gravity when a heat flux is imposed without convection. 
In equilibrium phase coexistence under gravity, a phase with a higher mass density is located below a phase with a lower mass density. Supposing that the phase with a higher mass density is preferable at low temperatures, such as a liquid phase in liquid-gas coexistence, a somewhat frustrating situation occurs when the heat flux is imposed against gravity. 
Therefore, it might be possible that the liquid floats over the gas against gravity.

In this Letter, we explore the phenomenon through molecular dynamics simulations. We find that when the directions of the gravitational force and the heat flux are opposite to each other, the liquid floats over the gas. No convection occurs, but a persistent heat flux generates an extra force balanced with gravity. The height of the floating liquid is characterized by a dimensionless parameter that represents the ratio of the temperature gradient to gravity. This scaling relation enables us to quantitatively predict the height of the floating liquid in a real experimental setup. Furthermore, we show that if a steady state is realized in the setup, there should be a region where the metastable states at equilibrium become stable under gravity and heat flow. In particular, the liquid floats only when the gas in the low-temperature region is supercooled at equilibrium.

{\em Setup.---} 
Liquid-gas transition occurs universally for any molecular system and even for noble gases without electric interactions. The noble gases have been modeled as simple systems using a Lennard--Jones potential, which is suitable for numerically studying various dynamic behaviors led by transitions \cite{holian1983, holian1988, kob1994, potoff1998, rosjorde2000,doliwa2003,ogushi2006,ishiyama2013,oh2013,watanabe2014,holyst2017,bresme2017,heinen2019,stephan2019,wen2021}. 
In our numerical simulations, we confine $N$ particles in a rectangular container with a height $L_x$ and side lengths $L_y$ and $L_z$, or height $L_x$ and side length $L_y$ in two-dimensional cases, under gravity.
For a collection of particle positions and momenta,
$\Gamma=(\bv{r}_1, \bv{r}_2, \cdots,\bv{r}_N, \bv{p}_1, \bv{p}_2, \cdots,\bv{p}_N)$, we assume the Hamiltonian as
\begin{equation}
H(\Gamma)=\sum_i\left[\frac{|\bv{p_i}|^2}{2m}+ \sum_{j<i} \phi(\abs{\bv{r}_i-\bv{r}_j}) + m g x_i + V_{\rm wall}(\bv{r}_i)\right],
\end{equation}  
where $m$ is the mass of the particles, and $g$ is the gravitational acceleration.
The two-body interaction potential $\phi$ is the 12-6 Lennard--Jones interaction 
\begin{align}
\phi(r)=4\vep\left[\left(\frac{\sigma}{r}\right)^{12}-\left(\frac{\sigma}{r}\right)^{6} \right]\theta(r_{c} - r). 
\label{e:LJ}
\end{align}
Here, $r$ is the distance between two particles, $\vep$ is the well depth, $\sigma$ is the particle diameter, $r_{c}$ is the cutoff length of
the interaction, and $\theta(r_{\rm c} - r)$ is the Heaviside step function. 
In the simulations, we set the cutoff length as $r_{\rm c} = 3\sigma$. 
We assume a fixed boundary condition at $x=0$ and $x=L_x$ 
using a soft-core repulsive wall represented by
$V_{\rm wall}(\bv{r}_i)$, where we adopt the Weeks--Chandler--Andersen potential 
to truncate the attracting interaction in \eqref{e:LJ} \cite{weeks1971}.
Other boundaries in the $y$ and $z$ directions are periodic.

The container is in contact with two heat baths at the top and bottom. To represent this setup, we perform molecular dynamics simulations
with Langevin thermostats having two temperatures $\HT$ and $\LT$. Each molecule evolves according to
\begin{align}
&\dot{\bv{p}}_i=-\pder{H}{\bv{r}_i}-\frac{\gamma(x_i)}{m} {\bv{p}}_i+\sqrt{2\gamma(x_i) \kB T_{\rm b}(x_i)}{\bv{\xi}_i}(t),\label{e:Langevin}
\end{align}
with $\dot{\bv{r}}_i={\bv{p}_i}/{m_i}$, where $\gamma(x_i)=1$ and $T_{\rm b}(x_i)=\HT$ in the region $0<x_i<8\sigma$, 
$\gamma(x_i)=1$ and $T_{\rm b}(x_i)=\LT$ in the region $L_x-8\sigma<x_i< L_x$, and $\gamma(x_i)=0$ in $8\sigma \le x_i \le L_x-8 \sigma$.
$\bv{\xi}_i(t)=(\xi_i^x(t),\xi_i^y(t),\xi_i^z(t))$ is Gaussian white noise that satisfies $\sbkt{\xi_i^a(t)}=0$ and $\sbkt{\xi_i^a(t)\xi_j^b(t')}=\delta_{i,j}\delta_{a,b}\delta(t-t')$, where $a$ and $b$ are $x$, $y$, or $z$.
We study the cases where $\HT$ and $\LT$ are far below the critical point and far above the crystallization temperature. 
The width of the thermostatted region, $8\sigma$, is sufficiently large compared to the mean free path in the liquid but comparable in the gas.
For later convenience, we define the middle temperature as $\Tm=(\HT+\LT)/2$ and the temperature difference as $\Delta T=\HT- \LT$.

\begin{figure}[t]
\centering
\includegraphics[width=0.35\textwidth]{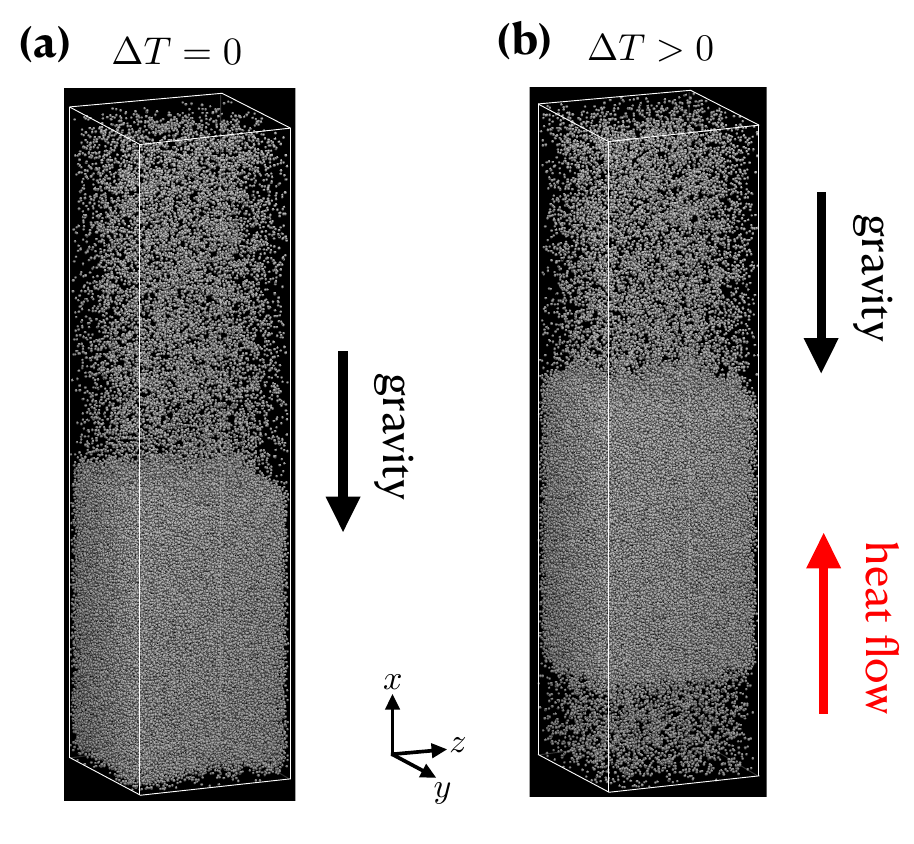}
\caption{Snapshots of the steady states under a gravitational force of $mg=6.25\times 10^{-5}\vep/\sigma$, $\kB \Tm =1.0\vep$, and mean density $N/L_{x}L_{y}L_{z}=0.3/\sigma^{3}$. (a) $\kB\Delta T=0$, (b) $\kB\Delta T = 0.1\vep$.
The aspect ratio is $L_{x}:L_y:L_z=4:1:1$, and the system size is $L_x=191\sigma$ corresponding to $N=131,072$. The width of each thermostatted region is $4.2$\% of $L_x$.
}
\label{f:snapshots}
\end{figure}

{\em Observation.---} 
We first prepared the equilibrium liquid-gas coexistence under gravity. 
We set the mean number density $N/(L_{x}L_{y}L_{z})$ and the temperature $\mT$
so that the volume ratio of the gas and liquid was almost 1. 
Figure \ref{f:snapshots} (a) shows
a typical configuration after sufficient relaxation 
for the system with the aspect ratio $L_x:L_y:L_z=4:1:1$ for $\Delta T=0$.
We see that the dense liquid is located in the lower region owing to gravity.

We then changed the values of $\HT$ and $\LT$ with $\kB\Delta T =0.1\vep$ while keeping the temperature $T$ at the middle value. 
Starting from the equilibrium state in Fig.~\ref{f:snapshots}(a), 
we find that the bulk of the liquid floats up gradually without separating into drops and
then keeps hovering as shown in a snapshot of the particle configuration in Fig. \ref{f:snapshots}(b). 
The relaxation process to a floating state is also shown in \cite{SM}.

\begin{figure}[t]
\centering
\includegraphics[width=0.4\textwidth]{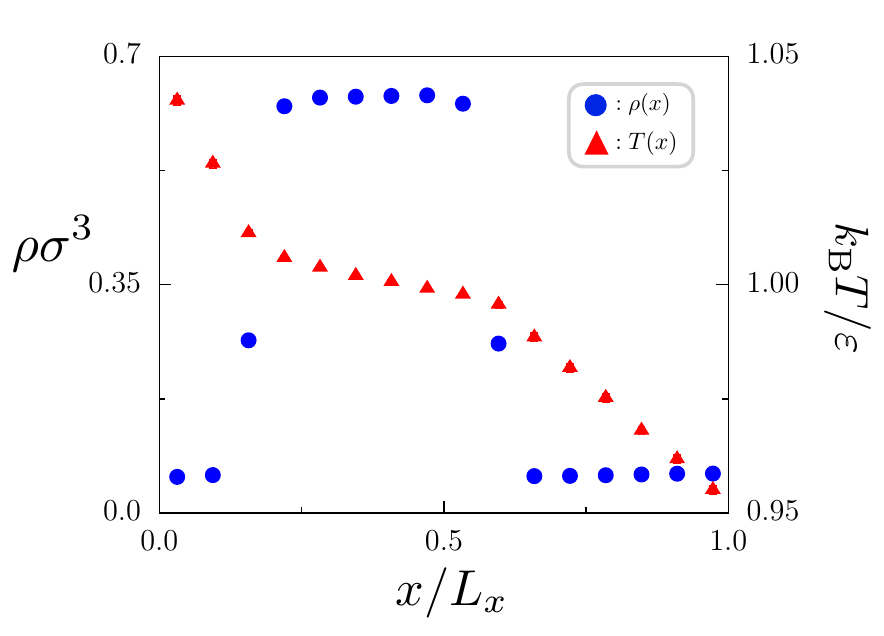}
\caption{Steady-state profiles of the number density $\rho(x)$ and temperature $\kinT$ for the system in Fig.~\ref{f:snapshots}(b) with $\chi=8.37$. The error bars are smaller than the symbols.  }
\label{f:profile_3d}
\end{figure}

To characterize the steady state, we calculated the number density per unit volume and the temperature profiles
\begin{align}
\rho(x)=\frac{\sum_i \langle \delta(x-x_i)\rangle}{L_yL_z},~
\kinT=\frac{\sum_i \langle\delta(x-x_i) |\bv{p}_i|^2\rangle}{3\kB L_yL_z m\rho(x)},\!
\end{align}
where $\langle\cdot\rangle$ is the long-term average after the relaxation.
In Fig.~\ref{f:profile_3d}, $\rho(x)$ shows two sharp interfaces separating the liquid and gas layers.
This means that the liquid layer is hovering and stationary. 
Correspondingly, $\kinT$ shows three regions with different slopes. 
The respective slopes result in a uniform heat current parallel to $x$.
The local velocities in the steady hovering state suggest that there is no convection in the hot-gas layer occupying the lower region \cite{SM}.
 
The liquid-hovering state is also observed in two-dimensional systems with $L_x:L_y=2:1$.
Some examples of the snapshot and the time evolution
in two dimensions are displayed in \cite{SM}.
Below, we concentrate on two-dimensional systems to examine the properties of the hovering state.

{\em Condition for hovering.---} 
We confirmed that the liquid hovers stationary under gravity and heat flow by varying the parameters of the container. The details of the following examples are demonstrated in \cite{SM}.
When the lateral boundary conditions are changed to be fixed, the floating liquid becomes slightly round owing to the repulsive interaction with the side walls.
The liquid continues to float without separating into pieces even when the container is horizontal with $L_x:L_y=1:2$.
However, when we impose wet boundary conditions using attractive interactions between each particle and the top or bottom walls, 
we observe that the liquid sticks to the top or bottom boundary,
or exhibits non-stationary motion.  

\begin{figure}[tb]
\centering
\includegraphics[width=0.45\textwidth]{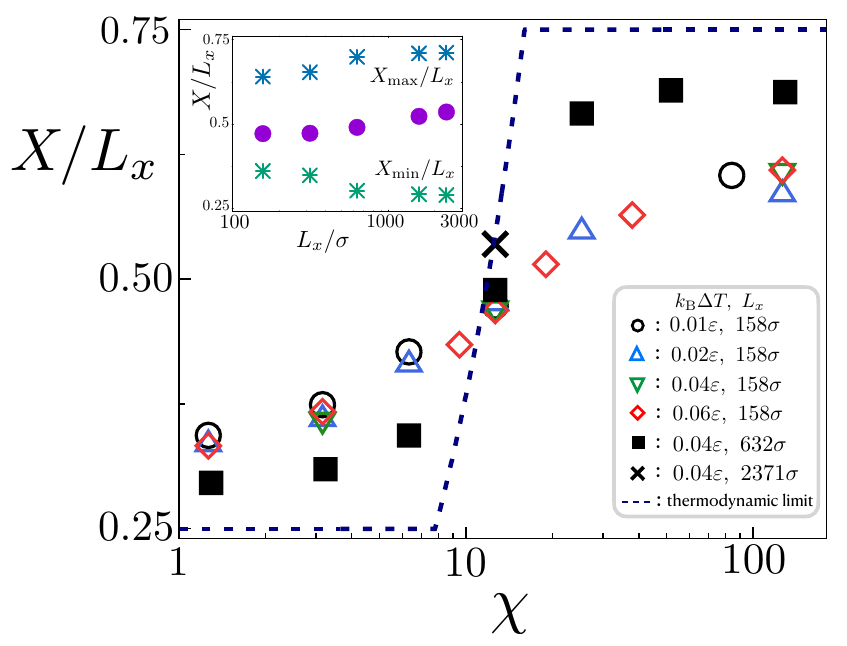}
\caption{
Center of mass $X/L_x$ versus $\chi=\kB\Delta T/mgL_x$.
Parameter values of $(\kB\Delta T, mg, L_x)$ for calculating each point are shown in \cite{SM}.
The error bars are smaller than the symbols.
The points for $L_x=2L_y=158\sigma$ (open symbols)
show the collapse $(g, \Delta T) \to (\alpha g, \alpha \Delta T)$ 
with four different degrees of nonequilibrium, $\kB\Delta T/\vep=0.01$, $0.02$, $0.04$, $0.06$ at $\kB \Tm/\vep=0.43$.
System size dependence of the collapsed curve is examined with $632\sigma$ (filled square), $2371\sigma$ (cross), and the thermodynamic limit with \eqref{e:chi-X} (dotted line).
(Inset) System size effects in $X/L_x$ and $X_{\max}/L_x$
at $\chi=12.65$ with $\kB\Delta T/\vep=0.04$. $L_x/\sigma=158$, $316$, $632$, $1581$, and $2371$. 
}
\label{f:scaling}
\end{figure}

We then focused on the original boundary condition mentioned in the
setup at the beginning. 
We took a mean density $N/(L_{x}L_{y}/\sigma^{2})=0.4$ and $\kB \Tm/\vep=0.43$
such that the volume ratio of the liquid and gas was almost 1.
The aspect ratio was fixed as $L_x:L_y=2:1$.
To characterize the hovering state, we examined how the center of mass 
\begin{align}
X=\sum_i \frac{\langle x_i\rangle}{N}
\end{align}
 depends on the temperature difference $\Delta T$ and the gravitational acceleration $g$
 and attempted to determine the functional form of $X(g, \Delta T)$ when $\Delta T>0$ and $g>0$. 
 We calculated $X$ for four values of $\kB\Delta T$ with changing $g$ 
for $L_x=158\sigma$ ($N=5.0\times 10^3$).
The important result is that $X(g, \Delta T)$ can be expressed in terms
of a scaling function. We first notice that the one-particle kinetic energy
difference between the top and the bottom is $k_B \Delta T$.
This quantity should be comparable with the potential
energy difference $mgL_x$. We then define the dimensionless parameter
\begin{align}
\chi\equiv\frac{\kB \Delta T}{mgL_{x}}.
\label{e:chi-def}
\end{align}
In Fig. ~\ref{f:scaling}, we plot $X(g, \Delta T)/L_x$ as a function of $\chi$. We find that
the data collapse on a single curve, $X=X(\chi)$. That is, the system is invariant
for the transformation of $(g, \Delta T) \to (\alpha g, \alpha \Delta T)
$ with any positive real $\alpha$. This result implies that
the temperature gradient plays the same role as the gravitational force.

To check the system size dependence, we examined $X(\chi)$ for $L_x=632\sigma$  ($N=8.0\times 10^4$) fixing $\kB\Delta T=0.04\vep$ and found the slight deviation of $X(\chi)$.
A point calculated in $L_x=2371\sigma$ ($N=1.125\times 10^6$) also deviates from the scaling function in $L_x=158\sigma$.
We then concentrated on the case $\chi=12.65$ and varied $L_x/\sigma$ from $158$ to $2372$, i.e., $5.0\times 10^3\le N\le 1.125\times 10^6$. See the inset of Fig.~\ref{f:scaling}.
Note that $X_{\min}<X<X_{\max}$, where $X_{\min}$ is the position of $X$ when $\Delta T=0$ for the respective value of $mg$, and $X_{\max}=L_x-X_{\min}$.
We observe a gradual increase in $X$ and $X_{\max}$ with $L_x$.
The scaling function tends to converge to the dotted line representing $X(\chi)$ in the thermodynamic limit, which will be derived below.

\begin{figure}[t]
\centering
\includegraphics[width=0.45\textwidth]{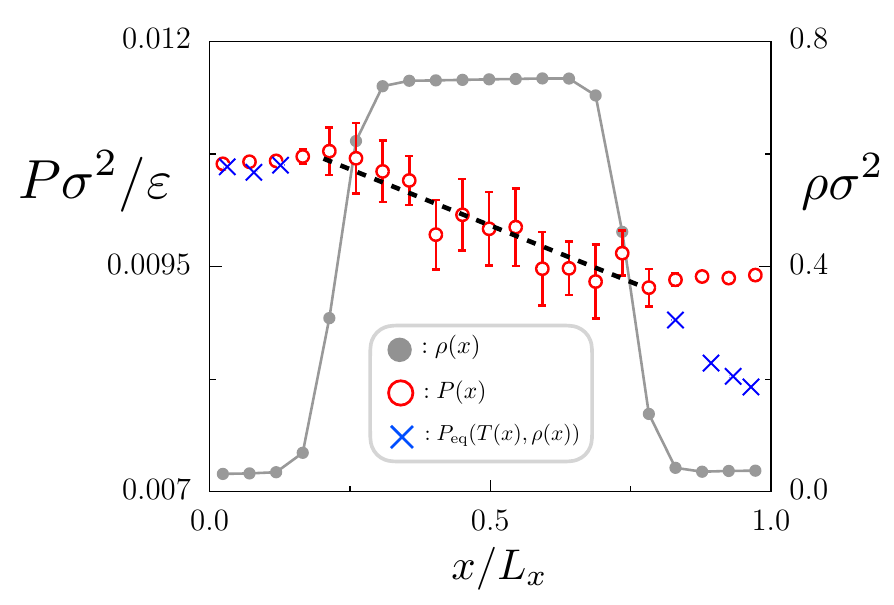}
\caption{Comparison of the local steady state and local equilibrium state with respect to pressure
for $\chi=12.65$, $\kB\Delta T/\vep=0.04$, $L_x = 632\sigma$, and $N=8.0\times 10^{4}$.
The error bars are twice the standard deviation.
The width of each thermostatted region is $1.3$\% of $L_x$.
The density profile $\rho(x)$ indicates the positions of the two interfaces.
The local steady pressure $P(x)$ is different from the local equilibrium pressure $P_\eq(\kinT,\rho(x))$ in the cold-gas layer.
}
\label{f:pxx_compare}
\end{figure}

{\em Thermodynamics.---} 
For the system of $L_x=632\sigma$ with $\chi=12.65$ and $\kB\Delta T/\vep=0.04$,
we first calculated $\rho(x)$, $\kinT$, and the Irving--Kirkwood stress tensor components $P_{xx}(x)$ and $P_{yy}(x)$ in the steady state \cite{irving1950}. 
Each profile is shown in \cite{SM}, where the consistency in the normal stress is shown as $P_{xx}(x)=P_{yy}(x)$ except for the vicinity of the interface. We thus define the local pressure as $P(x)\equiv P_{xx}(x)$.
In Fig. \ref{f:pxx_compare}, $\rho(x)$ and local pressure $P(x)$ are plotted simultaneously.
We find that $P(x)$ is almost constant in the hot and cold gas layers.

To examine the local equilibrium properties at each $x$,  
we numerically simulated the equilibrium system with the $NVT$ ensemble using the obtained local steady-state values $(\kinT, \rho(x))$. We set $N=8.0\times 10^4$ at $g=0$ with a periodic boundary condition for $x$ and $y$,
and calculated the virial pressure to determine the ``equilibrium'' pressure  $P_{\rm eq}(\kinT, \rho(x))$.
The details of the determination of $P_{\rm eq}(\kinT,\rho(x))$ are explained in \cite{SM}.
Figure ~\ref{f:pxx_compare} provides a simultaneous plot of $P(x)$ and $P_{\rm eq}(\kinT, \rho(x))$.
We find that $P(x) \simeq P_{\rm eq}(\kinT, \rho(x))$
in the hot-gas layer occupying the lower space, while $P(x) \not = P_{\rm eq}(\kinT, \rho(x))$
in the cold-gas layer occupying the upper space. 
The difference becomes larger with increasing distance from the interface between the liquid and the cold gas.
In the cold gas, the equilibrium state for $(\kinT,\rho(x))$ turns into the liquid-gas coexistence.
 Thus, the cold gas is considered to be supercooled, which is not equilibrium but metastable.

{\em Scaling function in the thermodynamic limit.--}
In Fig.~\ref{f:pxx_compare}, the two liquid-gas interfaces are in local equilibrium and therefore the pressures are saturated there. 
The profiles of $P(x)$ and $T(x)$ are piece-wise linear.
Based on these observations, the local pressure $P(x)$ in the liquid is identified as the saturation pressure $P_{\rm s}(T(x))$ for the local temperature $T(x)$ \cite{SM}.

Letting the position of the interfaces be $x^{\rm int}_1$ and $x^{\rm int}_2$ with $x^{\rm int}_1<x^{\rm int}_2$,
the force balance is written as
\begin{align}
P_{\rm s}(T(x^{\rm int}_1)))-P_{\rm s}(T(x^{\rm int}_2))=mg \rho^{\rm L}\Delta^{\rm L}
\label{e:balance}
\end{align}
with the width of the liquid layer $\Delta^{\rm L}\equiv x^{\rm int}_2-x^{\rm int}_1$
and  the number density $\rho^{\rm L}$ of the liquid.
Using $dP_{\rm s}/dx=(dP_{\rm s}/dT)(dT/dx)$, 
we extract the leading order contribution to (\ref{e:balance}) in the limit
$\Delta T/\Tm \to 0$. We then obtain
\begin{align}
\frac{|\nabla T|}{|\nabla T|^{\rm L}}=\frac{\chi}{\kB \rho^{\rm L}}\frac{dP_{\rm s}}{dT}
\label{e:chi-Gamma}
\end{align}
with the mean gradient $|\nabla T|=\Delta T/L_x$ and the gradient in the liquid 
$|\nabla T|^{\rm L}=-(T(x^{\rm int}_2)-T(x^{\rm int}_1))/\Delta^{\rm L}$,
where we have evaluated $dP_{\rm s}/dT$ at $T=\Tm$ \cite{SM}.

Since $T(x)$ is continuous and heat flux is uniform, $|\nabla T|/|\nabla T|^{\rm L}$ is linear in $X$ for $X_{\min}\le X\le X_{\max}$ when $\rho^{\rm L}\gg \rho^{\rm G}$. 
Then, the relation \eqref{e:chi-Gamma} indicates that $X$ is a linear function of $\chi$ in  $\chi_{\min}\le \chi\le\chi_{\max}$ \cite{SM}, where
\begin{align}
\chi_{\rm min/max}\!=\kB\rho^{\rm L}\!\left(\frac{dP_{\rm s}}{dT}\right)^{\! -1}\!\!\left[\frac{\Delta^{\rm L}}{L_x}+\frac{\kappa^{\rm L}}{\kappa_{2/1}^{\rm G}}\left(1\!-\!\frac{\Delta^{\rm L}}{L_x}\right)\right]\!.
\label{e:chi-X}
\end{align}
$\kappa^{\rm L}$, $\kappa^{\rm G}_1$ and $\kappa^{\rm G}_2$ are heat conductivities of the liquid, the hot gas, and the cold gas, respectively.
To be $\chi_{\rm min}<\chi_{\rm max}$, $\kappa_2^{\rm G}$ must be larger than $\kappa_1^{\rm G}$.
Consistently, $\kappa^{\rm G}_2/\kappa^{\rm G}_1$ is estimated about $2.5$ at the hovering state of $\chi=12.65$ and $\Delta^{\rm L}\simeq L_x/2$ for larger systems with $L_x/\sigma=632$ and $1581$ \cite{SM}.
Using numerical estimates for $\rho^{\rm L}$ and $dP_{\rm s}/dT$, the values of $\chi_{\rm min}$ and $\chi_{\rm max}$ are determined according to \eqref{e:chi-X}.
The obtained graph $X=X(\chi)$ is shown in Fig. \ref{f:scaling} as the dotted line, in which $\chi_{\rm min}=7.8$ and $\chi_{\rm max}=16$.

At last, we comment that the cold gas is thermodynamically unstable even in the macroscopic limit.
For the gas to be stable, the condition $P(x)\le P_{\rm s}(T(x))$ should be satisfied.
Combining the force balance $P(x)-P_{\rm s}(T(x^{\rm int}_2))=-mg\rho^{\rm G}(x-x^{\rm int}_2)$ in the cold gas with the Fourier law, we rewrite the stability condition as $- \kappa^\subL |\nabla T|^\subL (dP_{\rm s}/dT)\leq mg\rho^\subG \kappa^\subG_2$. We then eliminate $|\nabla T|^\subL$ by
substituting \eqref{e:chi-Gamma}. Finally, using the definition of $\chi$ in \eqref{e:chi-def}, we  obtain the stability condition for the cold gas as
\begin{align}
\rho^{\rm L}\kappa^{\rm L}\le \rho^{\rm G}\kappa_2^{\rm G}.
\label{e:stability}
\end{align}
This inequality is hardly reachable because $\rho^{\rm G} \ll \rho^{\rm L}$, and therefore, the cold gas is supercooled in general.

We here provide a physical explanation of how the cold gas
becomes metastable \cite{SM}.
Firstly, the pressure at the interface is determined as the saturation pressure $P_{\rm s}(T(x^{\rm int}_2))$, and the pressure above the interface is kept almost constant owing to the low density of the gas.
Secondly, according to the Fourier law, a small heat conductivity in gas results in a steep temperature gradient, and thus the upper region of the gas becomes colder than near the interface.
The combination of these two facts leads to the result that the gas situated sufficiently above the interface can only be supercooled.

{\em Summary and discussion.---}
We numerically observed the floating up of a liquid against gravity by imposing heat flow. The liquid hovers steadily without separating into pieces. 
The hovering state is characterized by the scaling function of the dimensionless parameter $\chi$. 
The cold gas situated above the liquid remains metastable and supercooled.  
The phenomenon has been explained from the thermodynamic
  argument based on the saturation property at the two liquid-gas interfaces.

The most important achievement will be experimental observation of these phenomena.
The scaling function $X(\chi)$ allows us to predict the temperature difference required for the floating up of the liquid against the gravity of the earth.
As an example, we consider the conditions under which xenon floats up.
The noble gas xenon shows liquid-gas coexistence around $220$ K, where the mass density ratio between the liquid and gas is $\rho^\subL/\rho^\subG =13$.
Applying $X(\chi)$ in Fig.~\ref{f:scaling}, the xenon in the container with $L_{x}=1\,$cm, $L_{y}=L_z=0.25\,$cm is expected to float up even with a small temperature difference of $\Delta T=0.2\,$K \cite{SM}. 
We thus believe that experimental observations are feasible. Not restricted to noble gases, familiar materials, such as nitrogens,  carbon dioxide, and water can show the phenomenon, too.
A possible difficulty is choosing a material for the container whose walls are repulsive to the fluid, such as the superhydrophobic materials used in the cold Leidenfrost examination for water \cite{quere2019}.
The effect of boundary properties, including details of the thermostats, on the hovering phenomenon should be investigated further.
Related to the experimental realization, a phenomenon that a slightly heavier phase is located above a lighter phase has been reported for liquid crystals in the heat conduction \cite{sakurai1999}.

At the end of this Letter, we briefly discuss convection in macroscopic systems.
The onset of thermal convection driven by the buoyancy force would be characterized by a threshold value of the Rayleigh number. 
Since the value of $\chi$ can be chosen independently of the Rayleigh number, the liquid hovering can occur in the absence of this type of convection.
One may consider another mechanism of convection driven by the temperature dependence of the surface tension, called Marangoni convection. We conjecture that this mechanism does not work in liquid-gas interfaces, because the temperature modulation along the interface leads to evaporation or condensation processes, which inhibit the flow caused by the surface tension.
Furthermore, even if convection occurs at a larger Rayleigh number than the threshold value, the hydrodynamic instability should be studied for the stationary hovering state when  $\chi \ge \chi_{\rm min}$. That is, the phenomena reported in this Letter provide a starting point for more complex states.

The authors thank A. Hisada, F. Kagawa, T. Nakamura,  S. Yukawa, K. Saito, and Y. Yamamura for useful discussions.
The numerical simulations were performed with LAMMPS on the supercomputer at ISSP at the University of Tokyo. 
The work of A.Y. was supported by JST and the Establishment of University Fellowships Towards the Creation of Science Technology Innovation under Grant Number JPMJFS2105.
This study was supported by JSPS KAKENHI Grant Numbers JP19KK0066, JP20K03765, JP19K03647, JP19H05795, JP20K20425, JP22H01144, and JP23K22415.

\clearpage

\onecolumngrid

\setcounter{figure}{0}
\setcounter{equation}{0}
\setcounter{table}{0}

\renewcommand{\theequation}{S.\arabic{equation}}
\renewcommand{\thefigure}{S.\arabic{figure}}
\renewcommand{\thetable}{S.\arabic{table}}

\begin{center}
{\large \bf Supplemental Material for  \protect \\ 
  `Heat-induced liquid hovering in liquid-gas coexistence under gravity' }\\
\vspace*{0.3cm}
Akira Yoshida$^{1}$, Naoko Nakagawa$^{1}$, and Shin-ichi Sasa$^{2}$
\\
\vspace*{0.1cm}
$^{1}${\small \it Department of Physics,
Ibaraki University, Mito 310-8512, Japan} \\
$^{2}${\small \it Department of Physics, Kyoto University,
Kyoto, 606-8502 Japan} 
\end{center}

The supplemental material consists of seven sections. In Sec. I, we show the relaxation process to the steady hovering state shown in Fig. 1. In Sec. II, we examine the robustness of the hovering states in two-dimensional systems. In Sec. III, we show the parameter values for calculating each point in Fig. 3 of the main text. Sections IV and V are for the definitions and the details of the quantities presented in Fig. 4. In Sec. VI, 
we derive Eqs. (7), (8), (9), and (10) and estimate $\chi_{\rm min}$ and $\chi_{\rm max}$ in the thermodynamic limit.
In Sec. VII, we quantitatively discuss experimental setups for observing the hovering state of noble gases.

\section{Relaxation process to the steady hovering state in Figs.1}\label{s:relaxation}

\begin{figure}[b]
\centering
\includegraphics[width=0.8\textwidth]{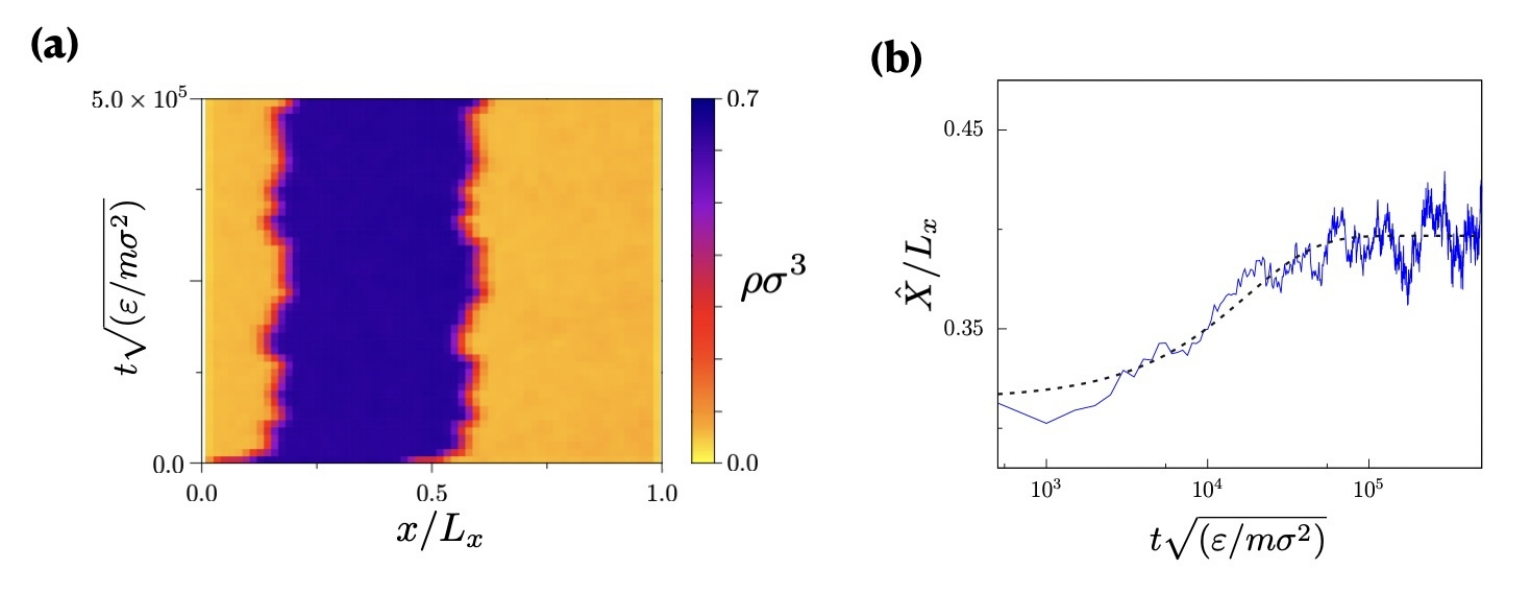}
\caption{
Relaxation process to the hovering state after imposing heat flow at $t=0$ for the three-dimensional system with $N=131,072$, $L_{x}:L_y:L_z=4:1:1$, $N/L_{x}L_{y}L_{z}=0.3/\sigma^{3}$,  $mg=6.25\times 10^{-5}\vep/\sigma$. 
(a) Dynamics of the density profile from the initial equilibrium state as shown in Fig. 1(a) to the steady hovering state in Fig. 1(b).
(b) Dynamics of the center of mass $\hat X$ as a logarithmic plot in time.  All parameters are equal to those used in Figs. 1. The dashed line is the result of fitting the data to the function $a-b\exp(-t/\tau)$, which leads to $\tau=1.75\times 10^4\sqrt{m\sigma^{2}/\vep}$.
}
\label{f:float_3d}
\end{figure}

We first explain the configuration displayed in Fig. 1(a) of the main text, which is prepared in equilibrium condition, $\kB \HT =\kB \LT=1.0\vep$. 
The configuration in Fig. 1(a) is obtained after the long calculation up to $t=1.75\times 10^4\sqrt{m\sigma^{2}/\vep}$  from an initial condition. The simulation time is a hundred times longer than the typical equilibration time.
We thus consider that the configuration in Fig. 1(a) belongs to equilibrium configurations.

Setting the system's configuration at $t=0$ as Fig. 1(a),
we perform the molecular dynamics simulation with $\kB\HT=1.05\vep$ and $\kB\LT=0.95 \vep$
corresponding to $\kB\Delta T= 0.1\vep$ and $\kB \Tm=1.0 \vep$. 
Figure \ref{f:float_3d} (a) shows the time evolution of the number density
profile 
\begin{align}
\rho(x)=\frac{1}{L_y L_z}\int_0^{L_y} dy \int_0^{L_z} dz~\rho(x,y,z)
\end{align}
for each $x$ as a color map.
It is observed that the liquid as a whole floats up against gravity and
then remains steadily hovering. 
Note that the liquid does not separate into pieces
while the liquid layer fluctuates.

To represent the relaxation dynamics shown in Fig. \ref{f:float_3d} (a),
we plot the time evolution of the center of mass,
\begin{align}
\hat X(t) \equiv \frac{1}{N}\sum_i x_i(t)
\end{align} 
in Fig. \ref{f:float_3d} (b).
This clearly shows an exponential relaxation fitted by 
$a-b\exp(-t/\tau)$, where $\tau=1.75\times 10^4\sqrt{m\sigma^{2}/\vep}$.  
We then conclude that the liquid hovers in the liquid-gas coexistence.
The snapshot in Fig. 1(b) is taken much after the relaxation time $\tau$, i.e.,
$t=3.0\times 10^6\sqrt{m\sigma^{2}/\vep}\simeq 170\tau$.

\section{Hovering in two dimensional systems}\label{s:hovering}

We examine the hovering of the liquid in two-dimensional systems with 
$N/L_x L_y=0.4/\sigma^2$ and $\kB \Tm=0.43 \vep$
by changing boundary conditions, the aspect ratio, and the wettability of the walls. 
In all cases, we fix the value of the scaling parameter $\chi$, defined by Eq. (6),  to 12.65 and the temperature difference $\kB \Delta T$ to $0.04 \vep$, and then the gravitational force is given by $mg=(\kB\Delta T/\vep)/(\chi L_x/\sigma)$.
Each initial condition is set as the corresponding equilibrium state prepared by another numerical simulation at $\Delta T=0$ with other parameters fixed.

Summarizing the following subsections, the hovering states are robust to the changes in the aspect ratio of the container and the boundary conditions.
We then expect that the floating up of the liquid is universal, although we need to choose the value of $\chi$ properly.
The dynamics of the liquid is affected by the wettability of the top and bottom walls.

\begin{figure}[b]
\centering
\includegraphics[width=0.95\textwidth]{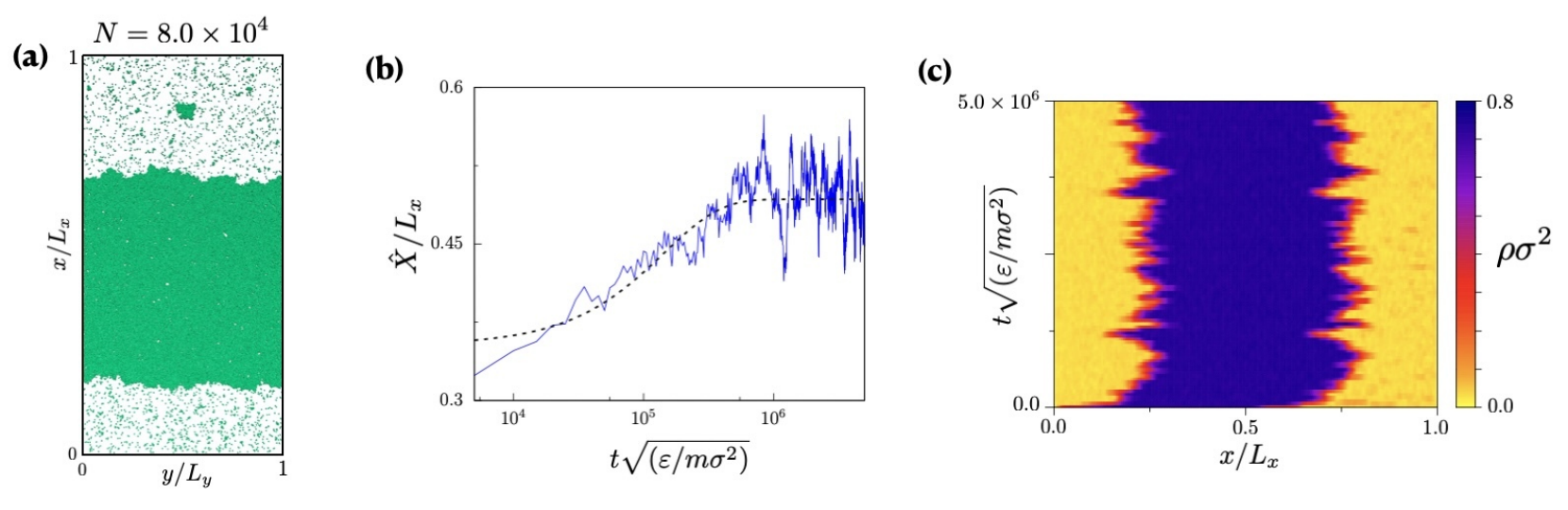}
\caption{Steady hovering state and relaxation process to the hovering state after imposing heat flow at $t=0$ for the two-dimensional system with periodic boundary conditions in the $y$-direction with $N=8.0\times 10^4$, $L_x/\sigma=632$, and $L_x:L_y=2:1$.
(a) Snapshot at $1.05\times 10^7 \sqrt{m\sigma^2/\vep}$ in the steady state. 
(b) Dynamics of the center of mass $\hat{X}$ as a logarithmic plot in time. The dashed line is the fitting curve as $a-b\exp(-t/\tau)$ where $\tau = 1.4 \times10^{5} \sqrt{m\sigma^2/\vep}$. 
(c) Dynamics of the density profile from the initial equilibrium state. 
(b) and (c) are plotted in $t<100\tau$}
\label{f:float_2d_80000}
\end{figure}

\begin{figure}[t]
\centering
\includegraphics[width=0.23\textwidth]{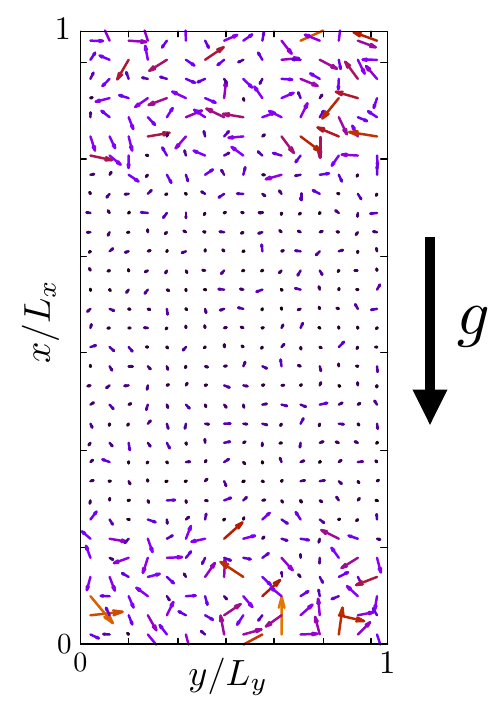}
\caption{Snapshot of the velocity field $\bv{v}(\bv{r},t)$ for $N=8.0\times 10^4$
at $t=1.0\times 10^7 \sqrt{m\sigma^{2}/\vep}$ corresponding to Fig. \ref{f:float_2d_80000}(a). 
The length of the arrows indicates the magnitude of the local velocity.}
 \label{f:convection}
\end{figure}

\subsection{Periodic boundary condition} \label{s:original}

First, we confine the particles in a box of an aspect ratio $L_x:L_y=2:1$
with fixed boundary conditions at $x=0$ and $x=L_x$ while periodic in the $y$-direction.
Figure \ref{f:float_2d_80000}(a) shows a snapshot at $t=1.05\times 10^7 \sqrt{m\sigma^{2}/\vep}$
for the system with $N=8.0\times 10^4$. As shown in Fig. \ref{f:float_2d_80000}(b),
the  relaxation time is estimated as $\tau=1.4\times 10^5 \sqrt{m\sigma^{2}/\vep}$.
The relaxation process to the hovering state is displayed in Fig.~\ref{f:float_2d_80000}(c)
as a color map of the local number density
$\rho(x)=\int_0^{L_y} dy~\rho(x,y)/L_y$.
It clearly shows that the liquid hovers steadily in the two-dimensional system, similar to the three-dimensional system.

\begin{figure}[b]
\centering
\begin{minipage}[h]{0.35\textwidth}
\centering
\includegraphics[width=1.0\textwidth]{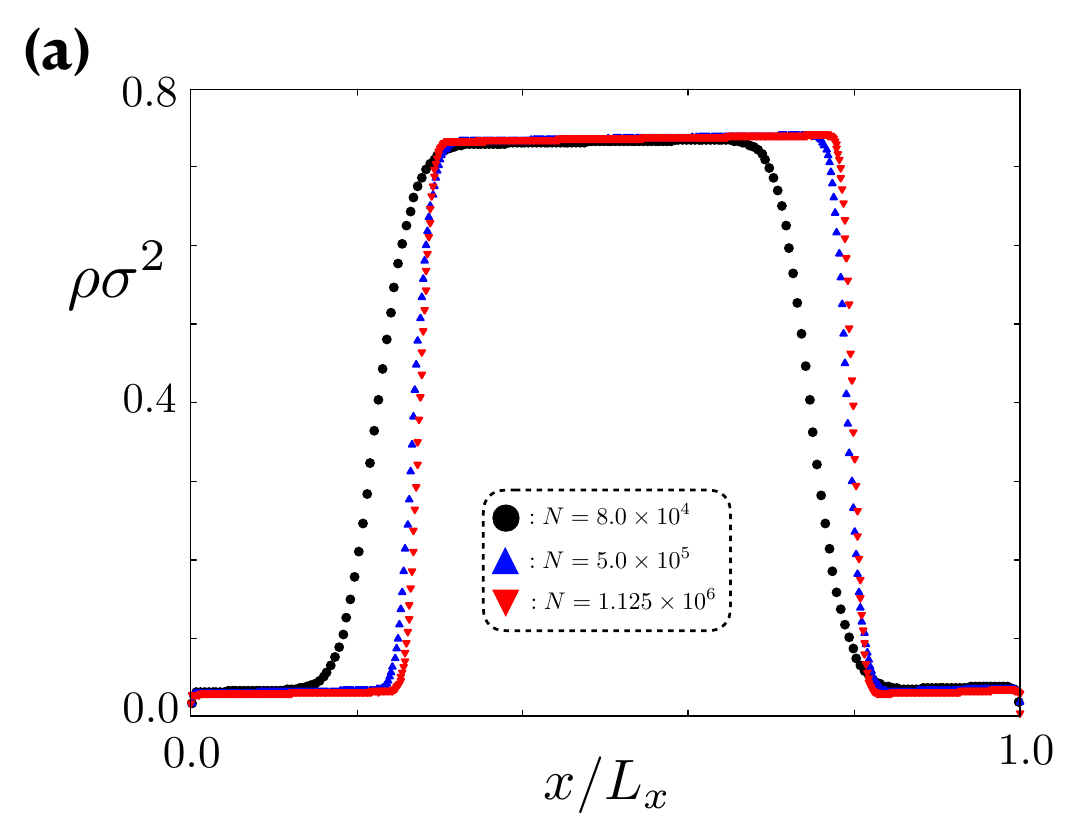}
\end{minipage} 
\begin{minipage}[h]{0.48\textwidth}
\centering
\includegraphics[width=0.98\textwidth]{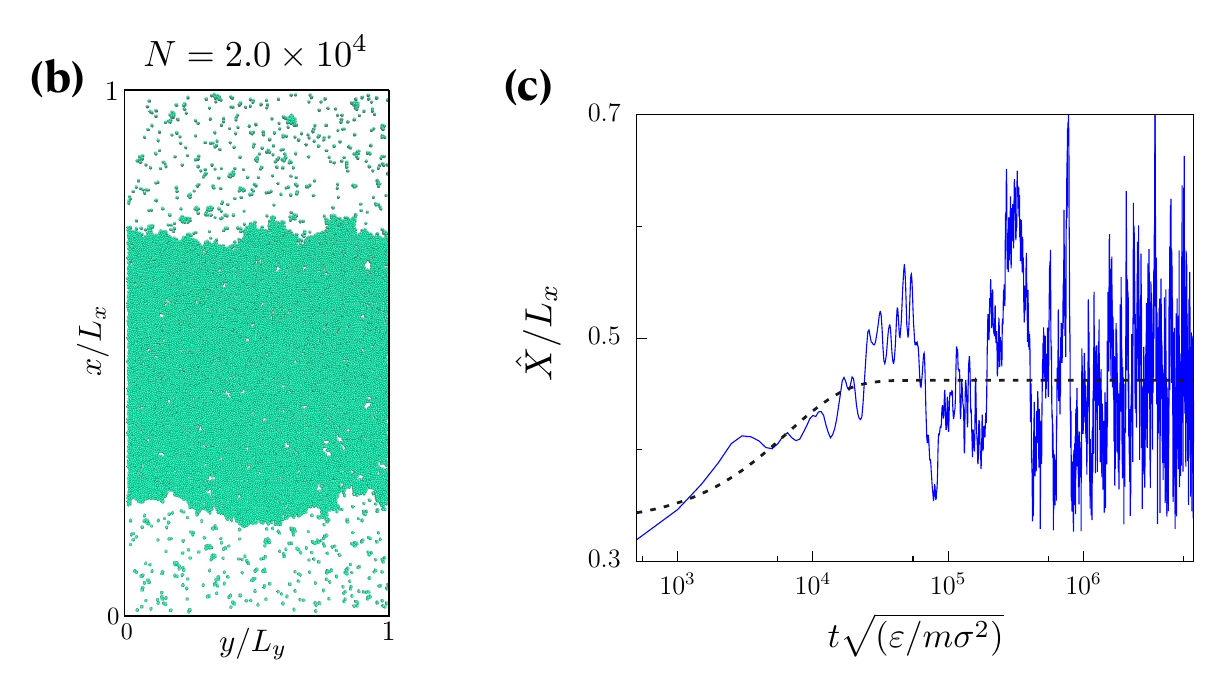}
\includegraphics[width=1.0\textwidth]{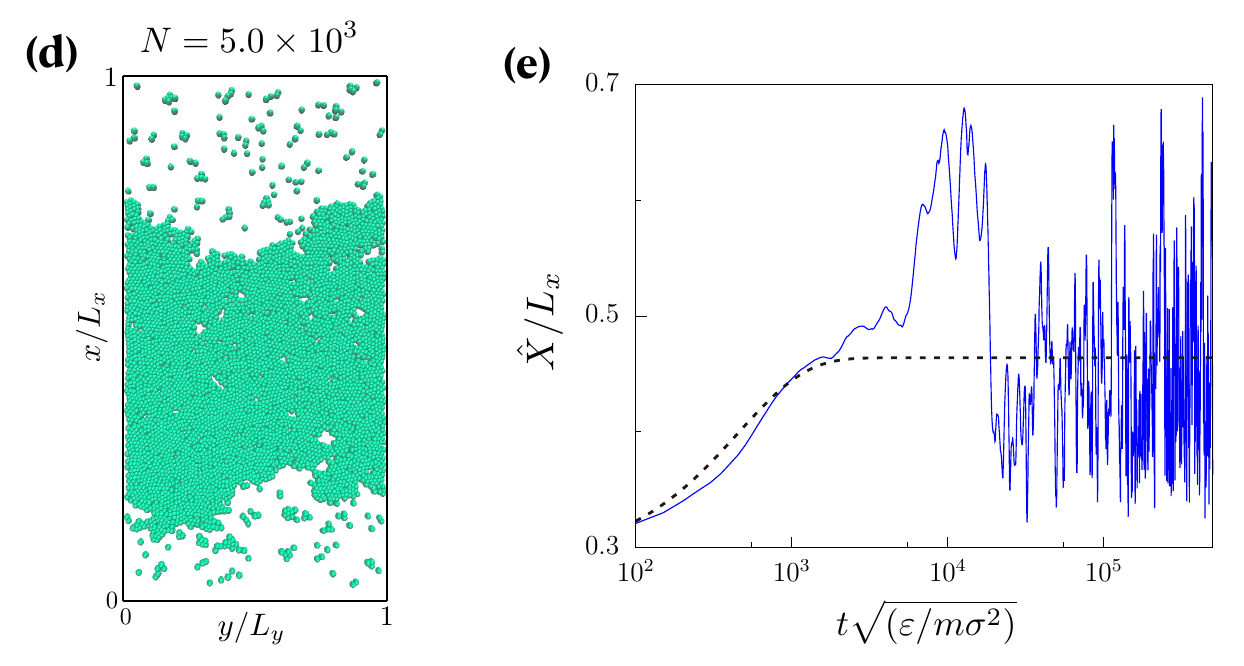}
\end{minipage}
\caption{
(a) Comparison of density profile $\rho(x)$ among three system sizes, $N=8.0\times 10^4$ with $L_x/\sigma=632$, $N=5.0\times 10^5$ with $L_x/\sigma=1581$, and $N=1.125\times 10^6$ with $L_x/\sigma=2371$.
(b) Snapshot of the steady hovering state for $N=2.0\times 10^4$ with $L_x/\sigma=316$ at $t=5.0 \times 10^6  \sqrt{m\sigma^{2}/\vep}$, taken 
sufficiently after the relaxation time $\tau$. 
The relaxation time $\tau$ after imposing heat flow is determined in (c) from the time evolution of $\hat X$.
(d) Snapshot of the steady hovering state for  $N=5.0\times 10^3$ with $L_x/\sigma=158$ at $t=4.5 \times 10^6  \sqrt{m\sigma^{2}/\vep}$, taken
sufficiently after the relaxation time $\tau$ determined from the time evolution of $\hat X$ in (e) after imposing heat flow.
The dashed lines in (c) and (e) are the fitting curves as $a-b\exp(-t/\tau)$. See text. (c) and (e) are plotted in $t<1000\tau$.
}
\label{f:float_2d_5000}
\end{figure}

Figure \ref{f:convection} depicts the snapshot of the velocity field defined as
\begin{align}
\bv{v}(\bv{r},t) \equiv \frac{\int^{t+\tau}_t ds \sum^{N}_{i=1} \bv{p}_i(s) \delta(\bv{r} - \bv{r}_i(s) )/m }{\int^{t+\tau}_t ds \sum^{N}_{i=1} \delta(\bv{r} - \bv{r}_i(s) ) },
\end{align}
where $\tau$ is chosen as the relaxation time for visibility of flow in steady states.
The velocity field shows a clear difference in the magnitude of velocity between the gas and liquid layers. 
The direction of the flow is obviously random and uncorrelated in space. 
The steady hovering of the liquid is not supported by the convection because there is no convective flow in either the hot or cold gas.
The absence of convective flow is further demonstrated by the movie of the velocity field in the Supplemental Material:\\
SMov: The dynamics of the velocity field $\bv{v}(\bv{r},t)$ for $1.0\times 10^7\sqrt{m\sigma^{2}/\vep} \leq t \leq 1.6\times 10^7\sqrt{m\sigma^{2}/\vep}$ in time width $\tau$.

The main text mentions that the hovering state is observed as the nonequilibrium steady state in various system sizes and is expected to survive in the thermodynamic limit. 
 Figure \ref{f:float_2d_5000}(a) shows the density profiles per unit volume for $N=8.0\times 10^4$, $N=5.0\times 10^5$, and $N=1.125\times 10^6$ with $L_x=632 \sigma$, $L_x=1581 \sigma$, and $L_x=2371 \sigma$, respectively, in a scaled space $x/L_x$.
We find that the interface profile becomes sharper and the position of the liquid layer seems to converge for large system sizes.
This behavior is consistent with that shown in the inset of Fig. 3 of the main text. 
The relaxation time to the hovering state increases rapidly with the system size.
As shown in Figs. \ref{f:float_2d_5000}(c) and (e), it is estimated as  $\tau= 4.5 \times 10^2 \sqrt{m\sigma^{2}/\vep}$ for $N=5.0\times 10^3$,
whereas  $\tau=6.5\times 10^3 \sqrt{m\sigma^{2}/\vep}$ for $N=2.0\times 10^4$. The relaxation time increases significantly with $N$, while the magnitude of the fluctuations decreases as seen in the long-time behaviors in Figs. \ref{f:float_2d_80000}(b), \ref{f:float_2d_5000}(b), and \ref{f:float_2d_5000}(d).

\subsection{Fixed boundary condition}

\begin{figure}[t]
\centering
\includegraphics[width=0.56\textwidth]{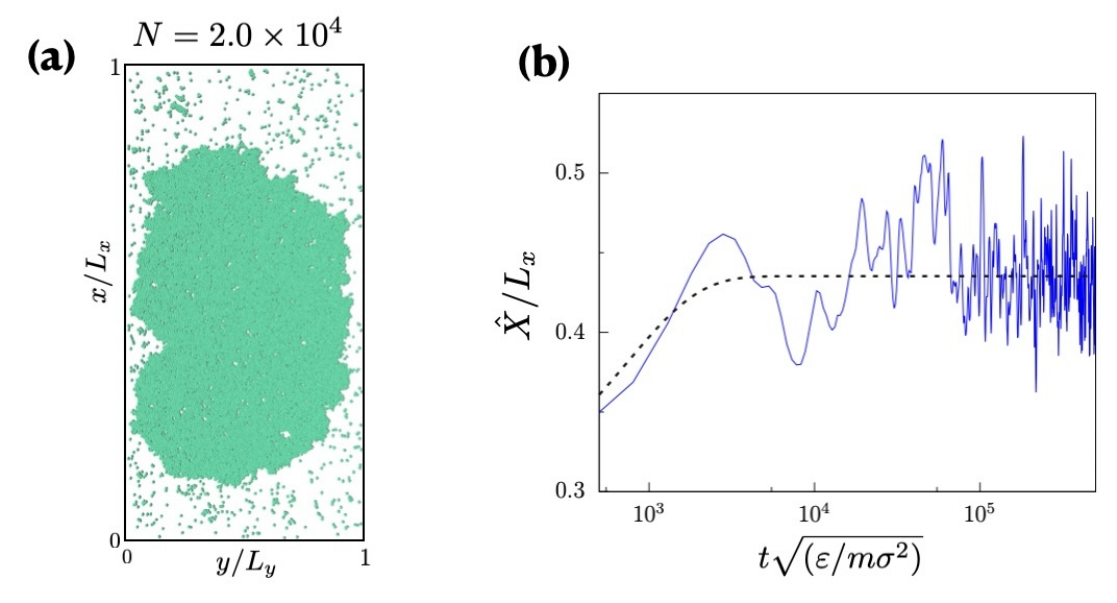}
\caption{
Steady hovering state and relaxation process to the hovering state under a fixed boundary condition in the $y$-direction. $N=2.0\times 10^4$ with $L_x/\sigma=316$ and $L_x:L_y=2:1$.
(a) Snapshot in the steady state. (b) Dynamics of the center of mass $\hat{X}$ as a logarithmic plot in $t<1000\tau$, where  $\tau = 7.5 \times 10^2 \sqrt{m\sigma^2/\vep}$. The dashed line is the fitting curve as $a-b\exp(-t/\tau)$.
}
\label{f:ywall_20000}
\end{figure}

We modify the boundary condition at $y=0$ and $y=L_y$
by replacing the periodic boundary condition with the fixed boundary condition.
Precisely, all four walls of the container are soft-core expressed by 
the  WCA potential. 
For the system with $N=2.0\times 10^4$ and $L_x:L_y=2:1$, the liquid floats up and hovers steadily after the relaxation.
The relaxation time $\tau$ is $7.5 \times 10^2 \sqrt{m\sigma^{2}/\vep}$ as examined in Fig.\ref{f:ywall_20000}(b), which is one order magnitude smaller than the relaxation time for the same value of $N$ with the periodic boundary condition in Fig.~\ref{f:float_2d_5000}(b).
Figure \ref{f:ywall_20000}(a)  is a configuration at $t=2.5\times 10^6\sqrt{m\sigma^2/\vep}$ sufficiently after the relaxation time.
Thin layers of low number density are found near the side walls. This causes the liquid to take on a rounded shape instead of the flat shape seen in the previous case (see Fig.~\ref{f:float_2d_80000}). This is due to the repulsive interaction with the lateral walls. Irrespective of the existence of the thin gas-like layers, the liquid continues to float in the center of the space.

\begin{figure}[b]
\centering
\includegraphics[width=0.7\textwidth]{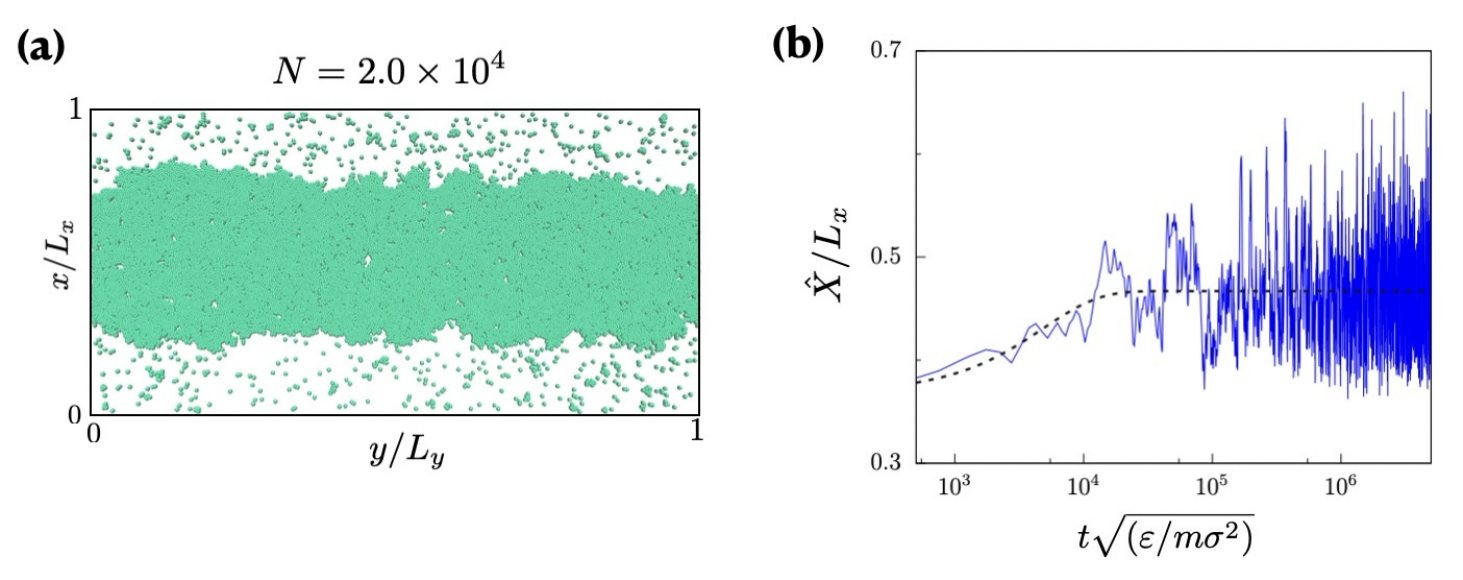}
\caption{Steady hovering state and relaxation process to the hovering state. $N=2.0\times 10^4$ with $L_x/\sigma=158$ and
 $L_x:L_y=1:2$. 
(a) Snapshot in the steady state. (b) Dynamics of the center of mass $\hat{X}$ as a logarithmic plot in $t<1000\tau$,
 where $\tau = 5.0\times 10^3 \sqrt{m\sigma^2/\vep}$. 
The dashed line is the fitting curve as $a-b\exp(-t/\tau)$.  }
\label{f:aspect_2_20000}
\end{figure}

\subsection{Different aspect ratio}

We change the aspect ratio to $L_x:L_y=1:2$ from 
 $L_x:L_y=2:1$. The boundary condition in the $y$-direction is periodic.
Figure \ref{f:aspect_2_20000}(a) displays the particle configuration for $N=2.0 \times 10^4$ at $t=5.0\times 10^6\sqrt{m\sigma^2/\vep}$ sufficiently after the relaxation.
The relaxation time is estimated as  $\tau = 5.0\times 10^3\sqrt{m\sigma^2/\vep}$ as shown in Fig. \ref{f:aspect_2_20000}(b), which is comparable to the relaxation time in the original aspect ratio $2:1$ for the same system size in Fig.\ref{f:float_2d_5000}(b).
The change in aspect ratio does not affect the stability of the hovering state or 
 the relaxation time $\tau$. 
We emphasize that the liquid still maintains a layered structure with two flat interfaces exceeding twice the side length $L_y$.

\subsection{Different wettability of walls}

\begin{figure}[t]
\centering
\includegraphics[width=0.6\textwidth]{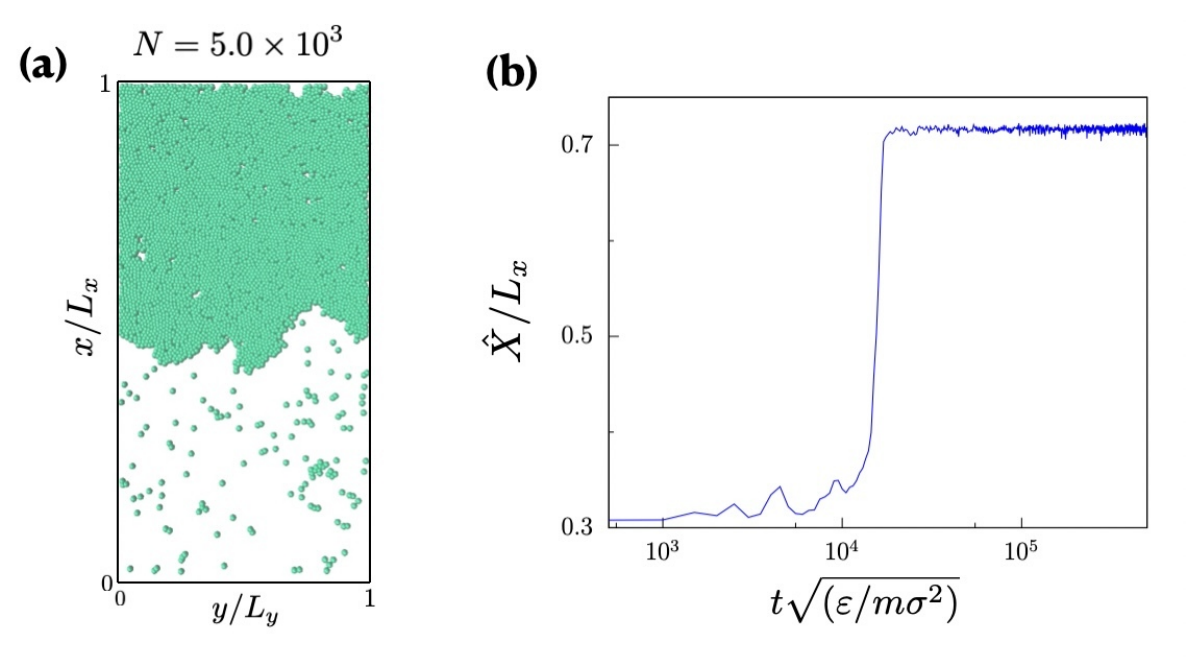}
\caption{Steady hovering state and relaxation process to the hovering state  in a container with attractive walls. $\chi=12.65$, N = $5.0\times 10^3$ with $L_x/\sigma=158$ and $L_x:L_y = 2 : 1$. 
(a) Snapshot at $t=2.5\times 10^6 \sqrt{m\sigma^2/\vep}$. (b) Dynamics of the center of mass $\hat{X}$ as a logarithmic plot in time. }
\label{f:wet_g0.00002}
\end{figure}

\begin{figure}[tb]
\centering
\includegraphics[width=0.7\textwidth]{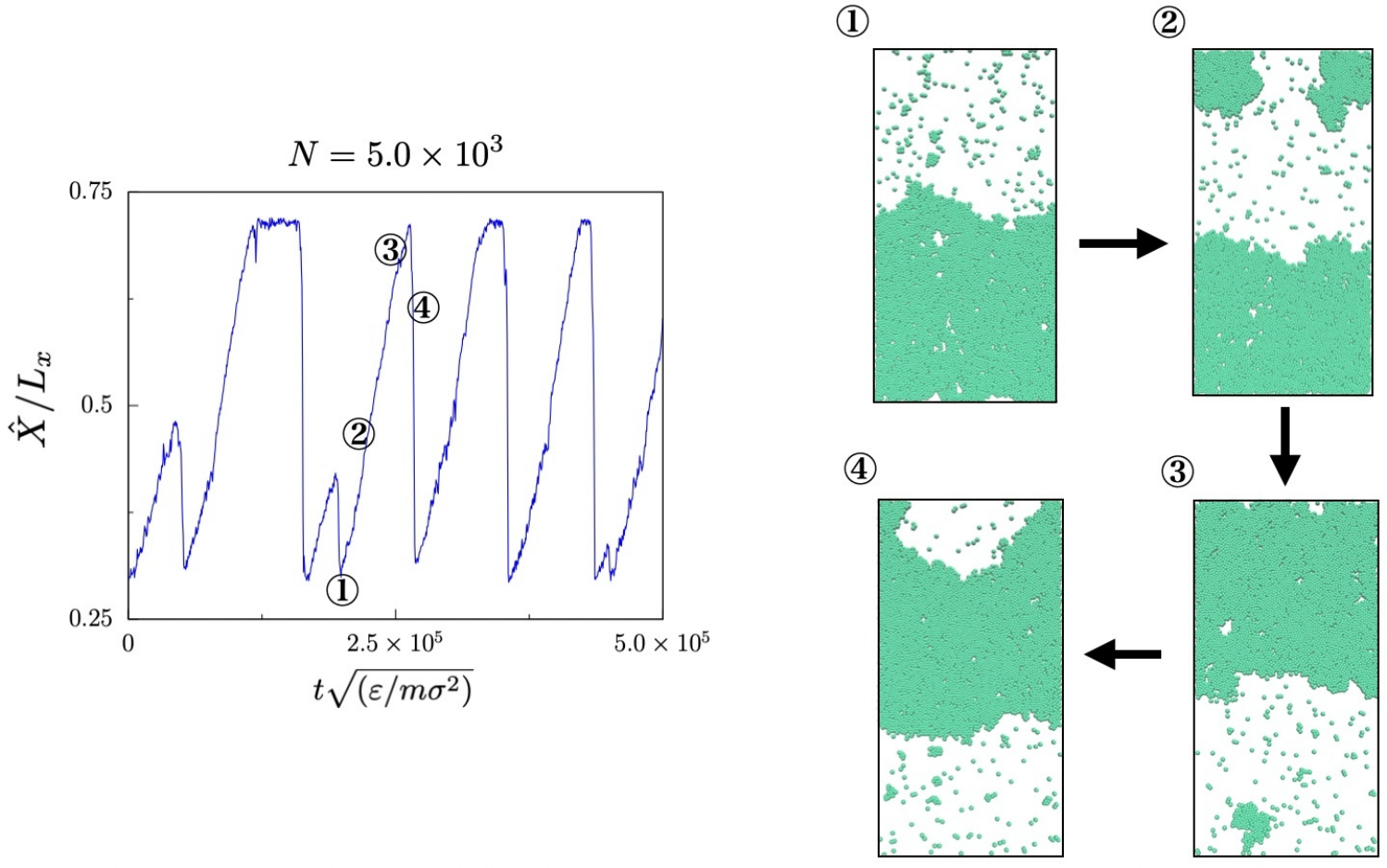}
\caption{Non-stationary dynamics in a container with attractive walls. $\chi=1.265$, $N =5.0\times 10^3$ with $L_x/\sigma=158$, and $L_x:L_y = 2 : 1$.  The value of $mg $ is ten times greater than that in the other simulations while keeping $\kB \Delta T /\vep=0.04$.
The liquid repeats intermittent dynamics. A typical time evolution is demonstrated by the sequential snapshots \textcircled{\scriptsize 1}, \textcircled{\scriptsize 2}, \textcircled{\scriptsize 3}, and \textcircled{\scriptsize 4}.
}
\label{f:wet_g0.0002}
\end{figure}

Finally, we change the interaction between each wall at $x=0$ or $x=L_x$ and each particle to the Lennard--Jones potential providing an attractive interaction. This change introduces the wettability of the walls.
We take $L_x:L_y=2:1$ and $N=5.0\times 10^3$ with the periodic boundary conditions in the $y$-direction.
We examine the steady states for $\chi=12.65$ and $\chi=1.265$.

The value $\chi=12.65$ corresponds to the common value used in this section. Figure \ref{f:wet_g0.00002}(a) shows a typical snapshot of the steady state and (b) illustrates the relaxation process. 
Comparing Fig. \ref{f:wet_g0.00002}(a) with Fig. \ref{f:float_2d_5000}(c), obtained from the same system except for the wettability of the walls, we notice that the liquid in Fig. \ref{f:wet_g0.00002}(a) occupies the upper region of the container. 
Note that the numerical examinations up to here with various dry wall systems show the common behavior that the liquid hovers in the middle of the container at $\chi=12.65$.
The robustness of the hovering state has been demonstrated by the scaling relation shown in Fig. 3 in the main text.
Nevertheless,  the wettability of the walls modifies significantly the degree of floatability
although it does not reduce the property that causes the liquid to float.

A remarkable difference appears in the dynamics. Fig. \ref{f:wet_g0.00002}(b) shows that the liquid sticks to the bottom until $t\sim 10^4\sqrt{m\sigma^2/\vep}$ and then suddenly floats up in a short time and sticks to the top wall. The time evolution in Fig. \ref{f:wet_g0.00002}(b) is not fitted by the exponential curve, which is qualitatively different from the exponential relaxation in Fig. \ref{f:float_2d_5000}(d). The comparison of the dynamics also shows a qualitative difference in the fluctuation after reaching the steady states. In addition, the time taken for the liquid to float up is ten times longer in Fig. \ref{f:wet_g0.00002}(b) than in Fig. \ref{f:float_2d_5000}(d), i.e., $t\sim 10^3\sqrt{m\sigma^2/\vep}$.

We then proceed to a strong gravity condition.
The value $\chi=1.265$ is one-tenth smaller than the previous common value and indicates that the gravity is relatively strong compared to the temperature difference. 
According to the scaling relation  in the dry wall systems shown in Fig. 3 of the main text, 
the liquid is hardly expected to float up at $\chi=1.265$.
Nevertheless, by replacing the walls with wet ones, the liquid floats up against the strong gravity.
The liquid unexpectedly falls to the bottom and repeatedly floats and falls as shown in Fig. \ref{f:wet_g0.0002}. 
Such intermittent behavior implies that the wettability of the walls introduces a qualitative change in the buoyancy property. These interesting dynamical problems remain as a future study.

\newpage

\section{Numerical parameters for Fig.~3 in the main text}

Each point of Fig.~3 in the main text was calculated by setting $\kB \Delta T/\vep$ and $mg\sigma/\vep$ as providing a given value of $\chi~(=\kB\Delta T/mgL_x)$.
 All specific values of $\kB \Delta T/\vep$ and $mg\sigma/\vep$ are described in Table \ref{t:parameterF3}.
For $L_x/\sigma=158$ $(N=5.0\times 10^3)$,  we calculated $10$ values of $\chi$, and for each value of $\chi$ we calculated at most four sets of $(\kB\Delta T/\vep, mg\sigma/\vep)$.
There are $7$ values of $\chi$ for $L_x/\sigma=632$ $(N=8.0\times 10^4)$, where we fix the value of $\kB\Delta T/\vep$ to $0.04$ at which the degree of nonequilibrium is $\Delta T/T_{\rm m}=0.093$.
For the largest system size of $L_x/\sigma=2371$ $(N=1.125\times 10^6)$, a single point at $\chi=12.65$ is calculated.

\begin{table}[hbt]
\centering
\begin{tabular}{| c | c | c | c | }
\hline
 ~~~~ $ L_x/\sigma $ ~~~~ &  ~~~~ $ \chi $ ~~~~~~~~  & ~~~~ $\kB \Delta T/\vep$ ~~~~  & ~~~~ $mg \sigma/\vep $ ~~~~ \\
\hline
\hline
  \multirow{24}{*}{158}  &  \multirow{3}{*}{1.265}   &  0.01 & $5.00\times 10^{-5}$ \\
 			&  	&0.02 &  $1.00\times 10^{-4}$ \\
			&  	& 0.06 &   $3.00\times 10^{-4}$  \\
\cline{2-4}
 		&  \multirow{5}{*}{3.162}  & 0.01 &  $2.00\times 10^{-5}$  \\
		&  	& 0.02 &   $4.00\times 10^{-5}$  \\
		&  	& 0.04 &   $8.00\times 10^{-5}$  \\
		&  	& 0.06 &   $1.20\times 10^{-4}$  \\
\cline{2-4}
 		&  \multirow{2}{*}{6.325}  & 0.01 & $1.00 \times 10^{-5}$   \\ 
		&  	& 0.02 &   $2.00 \times 10^{-5}$  \\
\cline{2-4}
 		& 9.486  & 0.06 & $4.00 \times 10^{-5}$   \\ 
\cline{2-4}
 		&  \multirow{5}{*}{12.65}  & 0.01 & $5.00 \times 10^{-6}$  \\ 
		&  	& 0.02 &   $1.00 \times 10^{-5}$  \\
		&  	& 0.04 &   $2.00 \times 10^{-5}$  \\
		&  	& 0.06 &   $3.00 \times 10^{-5}$  \\
\cline{2-4}
 		& 18.97 & 0.06 & $2.00 \times 10^{-5}$   \\ 
 \cline{2-4}
 		& 25.30 & 0.02 & $5.00 \times 10^{-6}$   \\ 
\cline{2-4}
 		& 37.95 & 0.06 & $1.00 \times 10^{-5}$   \\ 
\cline{2-4}
 		& 84.33 & 0.01 & $7.50 \times 10^{-7}$   \\ 
\cline{2-4}
 		& \multirow{3}{*}{12.65}  & 0.02 & $1.00 \times 10^{-6}$   \\ 
		&  	& 0.04 &   $2.00 \times 10^{-6}$  \\
		&  	& 0.06 &   $3.00 \times 10^{-6}$  \\
\hline
\hline
\multirow{8}{*}{632}  & 1.265 & 0.04 & $5.00\times 10^{-5}$ \\
\cline{2-4}
 		& 3.162 & 0.04 & $2.00 \times 10^{-5}$  \\ 
\cline{2-4}
 		& 6.325 & 0.04 & $1.00 \times 10^{-5}$  \\ 
\cline{2-4}
 		& 12.65 & 0.04 & $5.00 \times 10^{-6}$  \\ 
\cline{2-4}
 		& 25.30 & 0.04 & $2.50 \times 10^{-6}$  \\ 
\cline{2-4}
 		& 51.78 & 0.04 & $1.25 \times 10^{-6}$  \\ 
\cline{2-4}
 		& 126.5 & 0.04 & $5.00 \times 10^{-7}$  \\ 
 \hline
 \hline
 2371 & 12.65 & 0.04 & $1.33 \times 10^{-6}$ \\
 \hline
\end{tabular}
\caption{Parameter values set for the calculation of each point in Fig. 3. See the figure caption of Fig. 3 in the main text
for the explanation.
}
\label{t:parameterF3}
\end{table}

\newpage

\begin{figure}[t]
\centering
\includegraphics[width=0.9\textwidth]{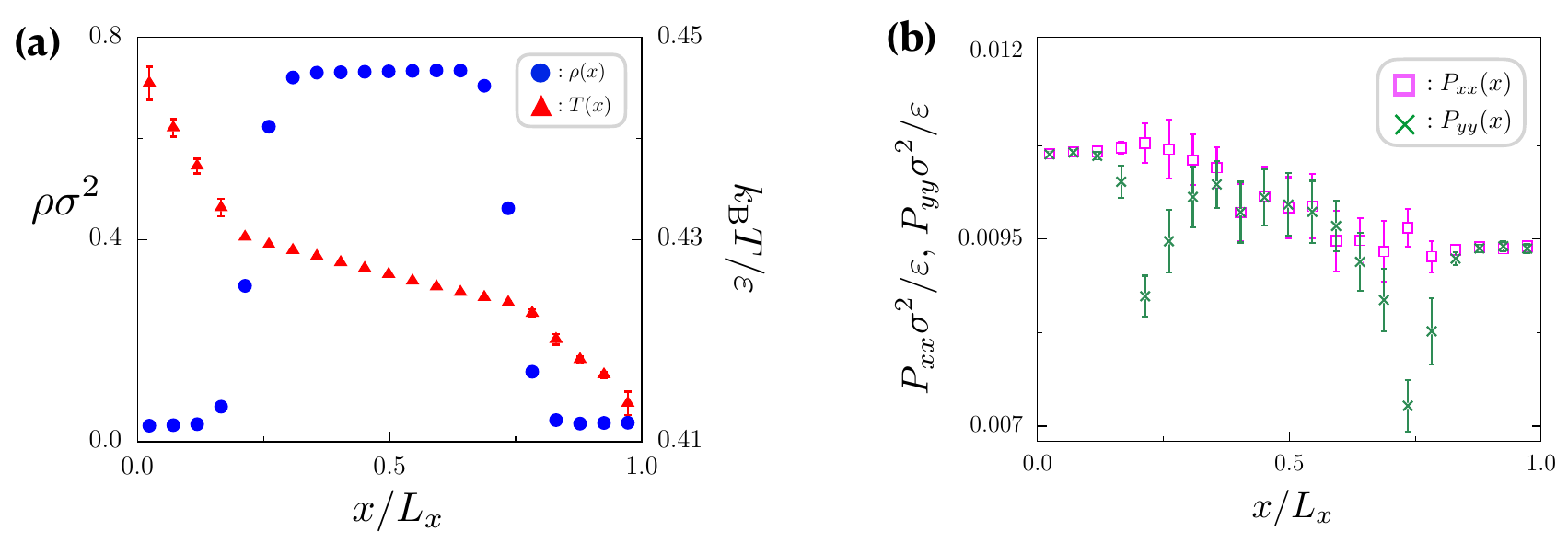}
\caption{Steady state profiles of the number density $\rho(x)$, temperature $T(x)$ and diagonal components of the Irving--Kirkwood stress tensor $P_{xx}(x)$ and $P_{yy}(x)$ for the system in Fig.~\ref{f:float_2d_80000}. The error bars are twice the standard deviation. }
\label{f:profile_2d}
\end{figure}

\section{Thermodynamic properties in steady states}\label{s:thermodynamic}

In Fig. 4 of the main text, $\rho(x)$ and $P(x)(=P_{xx}(x))$  are displayed, where $P_{xx}(x)$ and $P_{yy}(x)$ are defined as \cite{irving1950_}
\begin{align}
P_{xx}(x) = \left \langle \sum^{N}_{i=1} \left[ \frac{p^x_i p^x_i}{m} \delta(x-x_i) - \sum^{N}_{j>i}\frac{\partial \phi(r_{ij})}{\partial x_{ij}}(x_i - x_j) D(x_i,x_j;x) \right] \right \rangle ,\\
P_{yy}(x) = \left \langle \sum^{N}_{i=1} \left[  \frac{p^y_i p^y_i}{m} \delta(x-x_i) - \sum^{N}_{j>i}\frac{\partial \phi(r_{ij})}{\partial y_{ij}}(y_i - y_j) D(x_i,x_j;x) \right] \right \rangle 
\end{align}
with 
\begin{align}
D(x_i,x_j;x) = \int^{1}_{0} d \xi \delta(\xi x_i+(1-\xi)x_j-x) .
\end{align}
Here,  in this section,  we show the numerical data of $T(x)$ and $P_{yy}(x)$ in addition to $\rho(x)$ and $P_{xx}(x)$ shown in Fig. 4 of the main text.  
The temperature profile $T(x)$ is overlaid to $\rho(x)$  in Fig. \ref{f:profile_2d} (a) 
while the normal stress $P_{yy}(x)$ is plotted with the other normal stress $P_{xx}(x)$ in Fig. \ref{f:profile_2d} (b).
 It is observed that $P_{xx}(x)=P_{yy}(x)$ except for the interface regions. The discrepancy in the interface region is understood from the van der Waals stress that arises from the surface free energy.

\begin{table}[h]
\centering
\begin{tabular}{|c | c | c | c | c | c | c | c | c}
\hline
 index & a & b & c & d & e & f & g \\
 \hline
$\quad \kB T /\vep \quad$ & 0.444  & 0.440  & 0.437  & 0.420 & 0.417 &0.416  & 0.415 \\
\hline
$\rho/\sigma^2$ & \,0.0317\, &\, 0.0324\,  & \, 0.0337 \, & \, 0.0427 \, & \, 0.0351 \, & \, 0.0362 \, & \, 0.0374 \, \\
\hline\hline
$x/L_x$ & 0.03 & 0.08 & 0.13 & 0.83 & 0.89 & 0.93 & 0.96 \\
\hline
\end{tabular}
\caption{The values of $T(x)$ and $\rho(x)$ used in equilibrium simulations for each $x$.}
\label{t:T-rho}
\end{table}

\begin{figure}[h]
\centering
\includegraphics[width=0.45\textwidth]{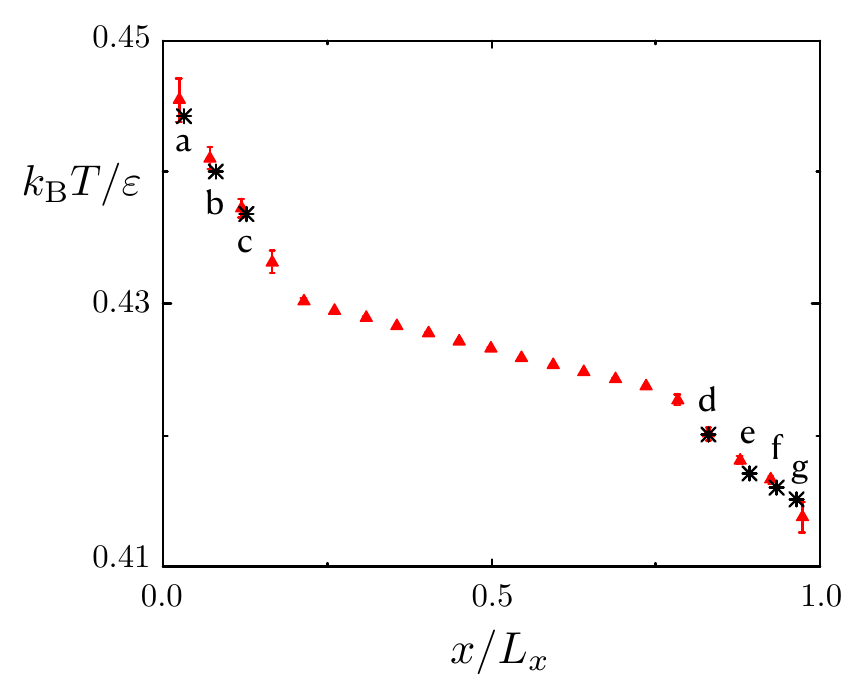}
\caption{
The temperatures of the seven equilibrium ensembles in Tab. \ref{t:T-rho} in comparison with the temperature profile shown in Fig. \ref{f:profile_2d}.
}
\label{f:compare_T}
\end{figure}

\section{Determination of $P_{\eq}(T(x),\rho(x))$ in Fig. 4}\label{s:Peq}

In this section, we explain the details for the determination of $P_{\eq}(T(x),\rho(x))$.

We set six $NVT$-ensembles, a, b, c, e, f, and g, whose temperature and number density are depicted in Table \ref{t:T-rho}.
As demonstrated in Fig. \ref{f:compare_T},
the values of $T$ and $\rho$ for each ensemble are chosen as they are along the profiles in Fig.~\ref{f:profile_2d}(a).
The ensembles a, b, and c are from the hot-gas layer while e, f, and g are from the cold-gas layer.
The ensemble d is in the vicinity of the interface.

We then perform the molecular dynamics simulations with the Nose--Hoover thermostat of the respective $T$
for the seven $NVT$-ensembles.
$N$ particles are confined in a rectangular box with the height $L_x$ and the side length $L_y$, so that the density is $N/L_x L_y = \rho$. We assume the periodic boundary conditions in the $x$ and $y$ directions and we fix $N=8.0\times 10^4$ and $L_x:L_y=2:1$. 
We calculate the instantaneous virial pressure at each moment as
\begin{align}
\hat{P} = \frac{1}{2L_x L_y} \left[ \sum^N_{i=1} \frac{|\bv{p}_i |^2}{m} - \sum^N_{i=1}\sum^N_{j>i} \frac{\partial \phi(r_{ij})}{\partial \bv{r}_{ij}} \cdot \bv{r}_{ij}  \right].
\end{align}
The equilibrium pressure is obtained from the long-term average of $\hat{P}$. 
The calculated virial pressure is considered as the equilibrium pressure for given $(T(x), \rho(x))$ of each ensemble, and then  denoted as $P_\eq(T(x),\rho(x))$, i.e., 
\begin{align}
\label{e:Peq}
P_{\eq}(T(x),\rho(x)) \equiv \bra \hat{P} \ket
\end{align}
using the respective value of $x$ in the Table \ref{t:T-rho}.

\begin{figure}[t]
\centering
\includegraphics[width=0.8\textwidth]{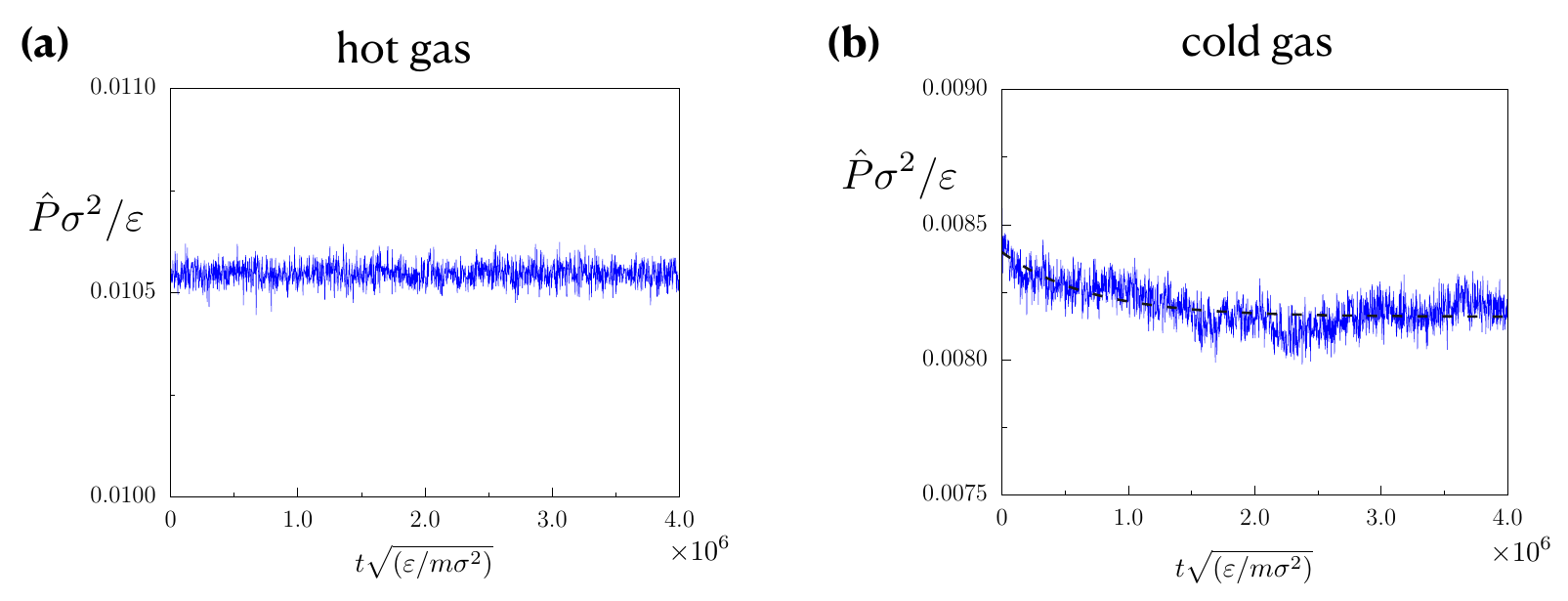}
\caption{Time evolutions of $\hat{P}$ for (a) the ensemble b and (b) the ensemble g.
The dashed line in (b) represents the fitting curve as $a-b\exp(-t/\tau)$ where $\tau=7.0\times 10^5 \sqrt{m\sigma^2/\vep}$. }
\label{f:p_dynamics}
\end{figure}

Figures \ref{f:p_dynamics} show the relaxation process of $\hat{P}$ for the ensembles b and g.
Here, the initial states at $t=0$ are set as the arrange of $16$ copies of the equilibrium configuration prepared by the same $NVT$-ensemble consisting of $N = 5,000$.
The ensemble b chosen from the hot-gas layer relaxes to equilibrium very quickly as shown in Fig. \ref{f:p_dynamics} (a).
In contrast, the ensemble g chosen from the cold-gas layer possesses a long relaxation time $\tau=7.0\times 10^5 \sqrt{m\sigma^2/\vep}$ determined from the exponential fitting as $a-b\exp(-t/\tau)$. See Fig. \ref{f:p_dynamics} (b). 
The relaxation time in the ensembles d, e, and f are obtained as comparable.
Thus, we calculated the time average for $t>2.5 \times 10^6 \sqrt{m\sigma^2/\vep}$.

\section{Details of the phenomenological argument}
\label{sec:pheno}

In Secs. \ref{press-liquid},  \ref{der-eq.8}, and \ref{der-eq.9}, we present details of the phenomenological arguments
for deriving Eqs. (7), (8), and (9) in the main text. 
We describe in Sec. \ref{s:numerical estimates} the details of the numerical estimate of $\chi_{\rm min}$ and $\chi_{\rm max}$ for drawing the dotted line in Fig. 3 of the main text.
In Secs. \ref{s:instability} and \ref{s:local-meta},  we show the derivation of Eq. (10) and demonstrate the thermodynamic instability of local states in the equilibrium phase diagram.

Below, let the positions of the liquid-gas interfaces be $x_1^{\rm int}$ and $x_2^{\rm int}$ satisfying $0\le x_1^{\rm int}\le x_2^{\rm int}\le L_x$ when the liquid hovers steadily.

\subsection{Saturation of the liquid}
\label{press-liquid}

We consider the local pressure $P(x)$ at a position $x$ inside the liquid layer, $x_1^{\rm int}\le x\le x_2^{\rm int}$.
Suppose that the pressures at the two liquid-gas interfaces are saturated.
That is, the local pressures are equal to the respective saturation
pressures,
\begin{align}
P(x_1^{\rm int})=P_{\rm s}(T(x_1^{\rm int})),\quad
P(x_2^{\rm int})=P_{\rm s}(T(x_2^{\rm int})).
\label{e:Ps-12}
\end{align}
Assuming a linear profile of $P(x)$ in $x$, we have
\begin{equation}
P(x)=\frac{P(x_2^{\rm int})(x-x_1^{\rm int})+ P(x_1^{\rm int})(x_2^{\rm int}-x)}
{x_2^{\rm int}-x_1^{\rm int}}.
\label{ps-0}
\end{equation}
We also have
\begin{align}
P_s(T(x)) &=P_s(T(x_1^{\rm int}))+\frac{dP_s}{dT}(T(x)-T(x_1^{\rm int}))
              \label{ps-1} , \\
          &=P_s(T(x_2^{\rm int}))+\frac{dP_s}{dT}(T(x)-T(x_2^{\rm int}))
              \label{ps-2}  .
\end{align}
Multiplying  \eqref{ps-1} by  $(x_2^{\rm int}-x)$,
\eqref{ps-2} by  $(x-x_1^{\rm int})$, and adding them, we obtain
\begin{equation}
  P_{\rm s}(T(x))=\frac{P_{\rm s}(T(x_2^{\rm int}))(x-x_1^{\rm int})
              + P_{\rm s}(T(x_1^{\rm int}))(x_2^{\rm int}-x)}
        {x_2^{\rm int}-x_1^{\rm int}},
\label{ps-4}  
\end{equation}
where we have used
\begin{equation}
  \frac{T(x)-T(x_1^{\rm int})}{x-x_1^{\rm int}}
=  
\frac{T(x_2^{\rm int})-T(x)}{x_2^{\rm int}-x}.
\end{equation}
Comparing \eqref{ps-0} and \eqref{ps-4}, and using  
\eqref{e:Ps-12}, we obtain 
\begin{align}
P(x)= P_{\rm s}(T(x)). 
\end{align}
We thus conclude that the whole of the liquid is saturated.

\subsection{Derivations of (7) and (8) in the main text}
\label{der-eq.8}

The difference of the pressure between the upper and lower interfaces,
which causes the buoyancy, balances with gravity as
\begin{align}
P(x_1^{\rm int})-P(x_2^{\rm int})=mg \rho^{\rm L} (x_2^{\rm int}-x_1^{\rm int}),
\label{e:buoy}
\end{align}
where $\rho^{\rm L}$ is the average number density of the liquid. 
Substituting \eqref{e:Ps-12} into \eqref{e:buoy} leads to (7) in the main text.

Using the magnitude of the temperature gradient in the liquid
\begin{align}
\left|\nabla T\right|^{\rm L}\equiv -\frac{T(x_2^{\rm int})-T(x_1^{\rm int})}{x_2^{\rm int}-x_1^{\rm int}},
\end{align}
and $\chi=\kB\Delta T/mgL_x$, 
(7)  is transformed into
\begin{align}
\frac{P_{\rm s}(T(x_2^{\rm int}))-P_{\rm s}(T(x_1^{\rm int}))}{T(x_2^{\rm int})-T(x_1^{\rm int})}
\left|\nabla T\right|^{\rm L}
=\frac{\kB \Delta T}{\chi L_x}\rho^{\rm L}.
\label{e:chi-Gamma-0}
\end{align}
Here, we extract the leading order contribution
  in the limit $\Delta T/\Tm \to 0$. By noting the smoothness of
  $P_{\rm s}(T)$ as a function of $T$ and the continuity of $T(x)$
  at the interface, we express the left side of \eqref{e:chi-Gamma-0} as
  $dP_{\rm s}/dT \left|\nabla T\right|^{\rm L}$ in this limit,
  where $dP_{\rm s}/dT$ is evaluated at $T=\Tm$. We thus obtain
\begin{align}
\frac{|\nabla T|}{|\nabla T|^{\rm L}}=\frac{\chi}{\kB \rho^{\rm L}}\frac{dP_{\rm s}}{dT},
\label{e:chi-Gamma-main}
\end{align}
which is (8) in the main text.
We note that \eqref{e:chi-Gamma-main} is also expressed by
\begin{align}
\frac{dP_{\rm s}}{dT} |\nabla T|^{\rm L}=mg \rho^{\rm L}.
\label{e:chi-Gamma-x}
\end{align}

\subsection{Derivation of (9) in the main text}
\label{der-eq.9}

Letting $\Delta^{\rm G}_1$, $\Delta^{\rm L}$, and $\Delta^{\rm G}_2$ be 
the widths of the hot-gas layer, the liquid layer, and the cold-gas layer, respectively,
\begin{align}
\Delta^{\rm G}_1=x_1^{\rm int},\quad
\Delta^{\rm L}=x_2^{\rm int}-x_1^{\rm int},\quad
\Delta^{\rm G}_1=L_x-x_2^{\rm int},
\label{e:dx-each}
\end{align}
the temperature differences in these layers are written as
\begin{align}
&T(x_1^{\rm int})-T(0)=-|\nabla T|^{\rm G}_1\Delta^{\rm G}_1,\\
&T(x_2^{\rm int})-T(x_1^{\rm int})=-|\nabla T|^{\rm L}\Delta^{\rm L},\\
&T(L_x)-T(x_2^{\rm int})=-|\nabla T|^{\rm G}_2\Delta^{\rm G}_2,
\end{align}
where $T(0)=\HT$ and $T(L_x)=\LT$. 
The total temperature difference, $\Delta T=\HT-\LT$, is connected to the temperature gradient in the three layers as
\begin{align}
\Delta T=|\nabla T|^{\rm G}_1\Delta^{\rm G}_1+|\nabla T|^{\rm L}\Delta^{\rm L}+|\nabla T|^{\rm G}_2\Delta^{\rm G}_2.
\label{e:dT-sum}
\end{align}
In the steady state, the heat flows in parallel to the $x$-axis and
the heat flux $J$ is constant in $x$.
Since $J>0$ for $\LT<\HT$, we have
\begin{align}
J=
\kappa_1^{\rm G} |\nabla T|^{\rm G}_1
=
\kappa^{\rm L} |\nabla T|^{\rm L}
=
\kappa_2^{\rm G} |\nabla T|^{\rm G}_2,
\label{e:J-uniform}
\end{align}
where $\kappa^{\rm L}$, $\kappa^{\rm G}_1$ and $\kappa^{\rm G}_2$ are heat conductivities of the liquid, the hot gas and the cold gas, respectively.
Using the equalities in \eqref{e:J-uniform},  the relation \eqref{e:dT-sum} leads to
\begin{align}
\frac{|\nabla T|}{|\nabla T|^{\rm L}}
=
\frac{\kappa^{\rm L}}{\kappa_1^{\rm G}}\frac{\Delta_1^{\rm G}}{L_x}
+\frac{\Delta^{\rm L}}{L_x}
+\frac{\kappa^{\rm L}}{\kappa_2^{\rm G}}\frac{\Delta_2^{\rm G}}{L_x}.
\label{e:GammaL-pre}
\end{align}

The center of mass $X$ can be formulated with
the number densities,  $\rho^{\rm L}$, $\rho^{\rm G}_1$, $\rho^{\rm G}_2$,
and the widths, $\Delta^{\rm L}$, $\Delta^{\rm G}_1$, $\Delta^{\rm G}_2$.
Especially when $\rho^{\rm L}\gg\rho_{1}^{\rm G}\simeq \rho_{2}^{\rm G}$,
the center of mass for the system is approximately given by
the center of mass for the liquid so that $X=\Delta_1^{\rm G}+{\Delta^{\rm L}}/{2}$, $X_{\rm min}=\Delta^{\rm L}/2$, and $X_{\rm max}=L_x-\Delta^{\rm L}/2$.
We then express $\Delta^{\rm G}_1$ and $\Delta^{\rm G}_2$ as
\begin{align}
\Delta_1^{\rm G}=X-\frac{\Delta^{\rm L}}{2}, \quad
\Delta_2^{\rm G}=L_x-X-\frac{\Delta^{\rm L}}{2}.
\label{e:Delta-X}
\end{align}
Substituting \eqref{e:Delta-X} into \eqref{e:GammaL-pre}, we have
\begin{align}
\frac{|\nabla T|}{|\nabla T|^{\rm L}}
=
\left(
\frac{\kappa^{\rm L}}{\kappa_1^{\rm G}}-\frac{\kappa^{\rm L}}{\kappa_2^{\rm G}}
\right)\frac{X}{L_x}
+\frac{\kappa^{\rm L}}{\kappa_2^{\rm G}}
+\frac{\Delta^{\rm L}}{L_x}
-\frac{\Delta^{\rm L}}{2L_x}\left(
\frac{\kappa^{\rm L}}{\kappa_1^{\rm G}}+\frac{\kappa^{\rm L}}{\kappa_2^{\rm G}}
\right),
\label{e:linear-X}
\end{align}
which is a linear function of $X$ in $X_{\rm min}\le X \le X_{\rm max}$ as mentioned in the main text.

Combining \eqref{e:linear-X} with \eqref{e:chi-Gamma-main},
we conclude that $X$ is  a linear function of $\chi$
in $\chi_{\rm min}\le \chi \le \chi_{\rm max}$.
When the liquid is on the bottom,  $X=X_{\rm min}$, $\Delta_1^{\rm G}=0$ and $\Delta_2^{\rm G}=L_x-\Delta^{\rm L}$, and therefore, 
\eqref{e:GammaL-pre} provides
\begin{align}
\frac{|\nabla T|}{|\nabla T|^{\rm L}}
=
\frac{\Delta^{\rm L}}{L_x}
+\frac{\kappa^{\rm L}}{\kappa_2^{\rm G}}\left(1-\frac{\Delta^{\rm L}}{L_x}\right)
\quad \mathrm{for} ~~{X=X_{\rm min}}.
\end{align}
Substituting this into \eqref{e:chi-Gamma-main}, we have
\begin{align}
  \chi_{\rm min}=\kB\rho^{\rm L}\left(\frac{dP_{\rm s}}{dT}\right)^{-1}\left[\frac{\Delta^{\rm L}}{L_x}+\frac{\kappa^{\rm L}}{\kappa_{2}^{\rm G}}\left(1-\frac{\Delta^{\rm L}}{L_x}\right)\right].
  \label{s30}
\end{align}
When the liquid is on the top of the container, $X=X_{\rm max}$, $\Delta_2^{\rm G}=0$ and $\Delta_1^{\rm G}=L_x-\Delta^{\rm L}$. We then have
\begin{align}
  \chi_{\rm max}=\kB\rho^{\rm L}\left(\frac{dP_{\rm s}}{dT}\right)^{-1}\left[\frac{\Delta^{\rm L}}{L_x}+\frac{\kappa^{\rm L}}{\kappa_{1}^{\rm G}}\left(1-\frac{\Delta^{\rm L}}{L_x}\right)\right].
  \label{s31}
\end{align}
Equations \eqref{s30} and \eqref{s31} are written as (9) in the main text.

\subsection{Numerical estimate of $\chi_{\rm min}$ and $\chi_{\rm max}$}
\label{s:numerical estimates}

\begin{figure}[b]
\centering
\includegraphics[width=0.4\textwidth]{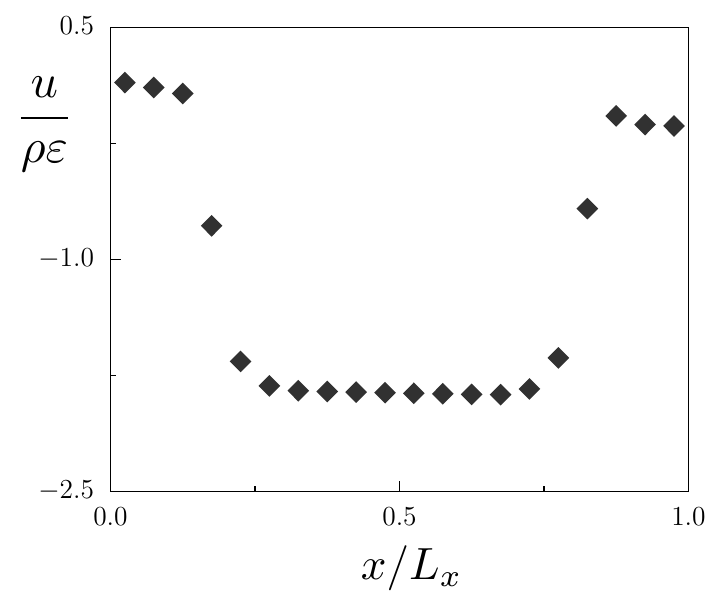}
\caption{Profile of the specific internal energy $\hat u(x)=u(x)/\rho(x)$ for the hovering state at $\chi=12.65$ in $L_x/\sigma=632$ ($N=8.0\times 10^4$).
$u(x)$ and $\rho(x)$ are the internal energy density  and the number density, respectively.
}
\label{f:u-profile}
\end{figure}

\begin{figure}[b]
\centering
\includegraphics[width=0.8\textwidth]{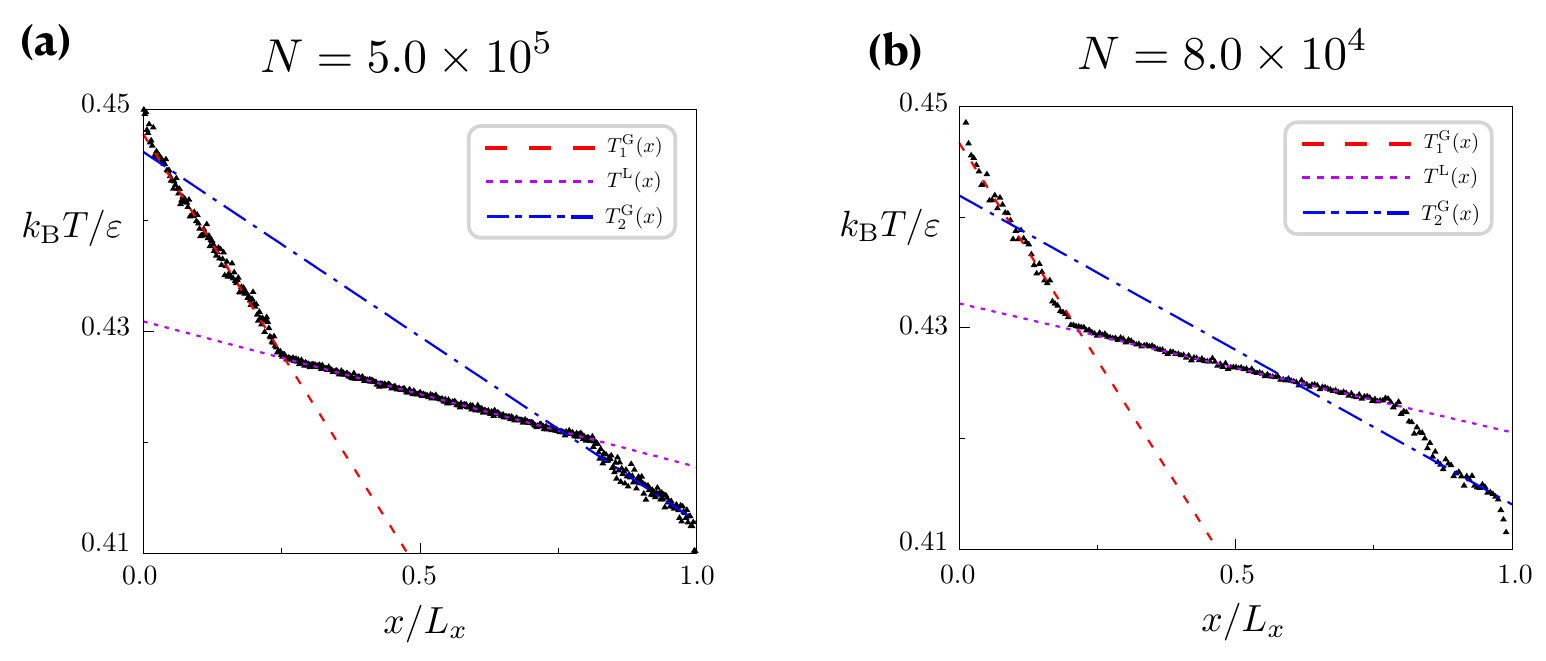}
\caption{Temperature profiles for the hovering state at $\chi=12.65$.  
(a) $L_x/\sigma=1581$ with $N=5.0\times 10^5$ and
(b) $L_x/\sigma=632$ with $N=8.0\times 10^4$.
The profiles are fitted by piece-wise linear function as
(a) $\kB T_1^{\rm G}(x)/\vep=-5.0\times 10^{-5} x+$const,  $\kB T_2^{\rm G}(x)/\vep=-2.1\times 10^{-5} x+$const, 
and  $\kB T^{\rm L}(x)/\vep=-8.3\times 10^{-5} x+$const,
and (b) $\kB T_1^{\rm G}(x)/\vep=-1.2\times 10^{-4} x+$const,  $\kB T_2^{\rm G}(x)/\vep=-4.4\times 10^{-5} x+$const, 
and  $\kB T^{\rm L}(x)/\vep=-1.8\times 10^{-5} x+$const.
A region near the liquid-gas interface is affected by the fluctuation of the position of the interface. We have avoided the region from the fitting range. Such a region becomes vanishing in sufficiently large systems as implied by the comparison between (a) and (b). 
}
\label{f:conductivity}
\end{figure}

\begin{figure}[t]
\centering
\includegraphics[width=0.5\textwidth]{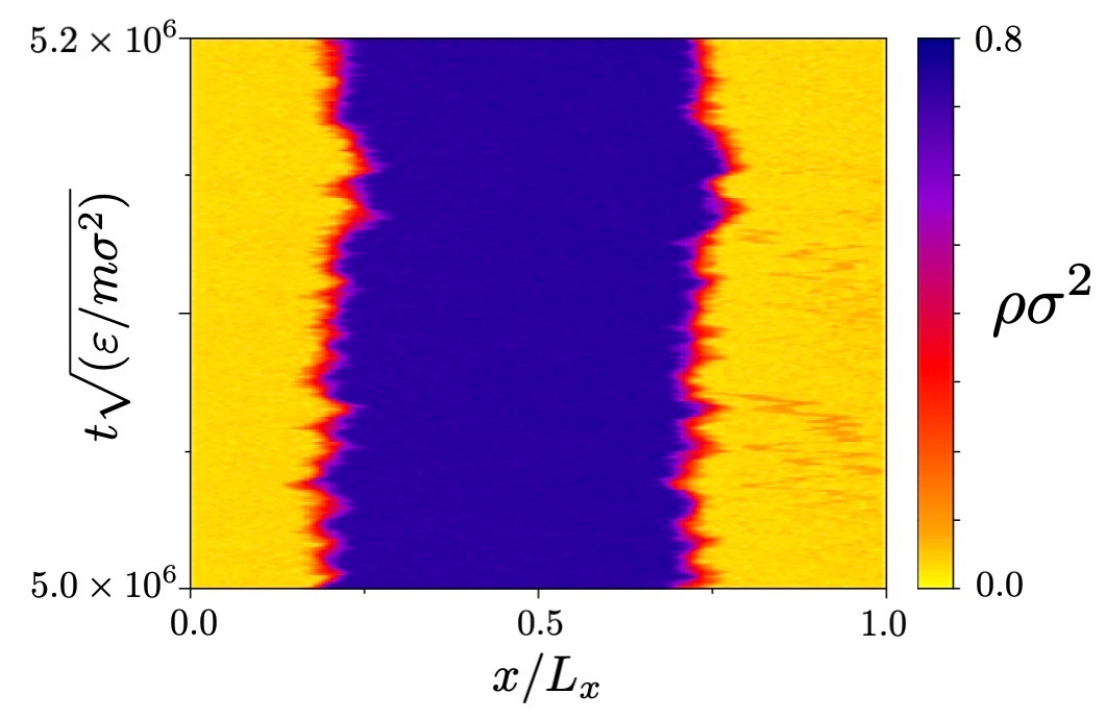}
\caption{Fluctuations of the hovering state at $\chi=12.65$ for the two-dimensional system with $L_x=632\sigma$ and $N=8.0\times 10^4$. 
Dynamics of the density profile is shown as a color map.
The left and right sides are both gas layers. In the cold gas layer on the right side, a few orange fluctuating lines continue from around the top of the gas to the interface with the liquid. The widths of the lines are comparable to the sizes of the liquid droplets. In the hot gas on the left side, the orange fluctuating lines hardly appear.
}
\label{f:droplet}
\end{figure}

For estimating $\chi_{\min}$ and $\chi_{\rm max}$ from (9), we need to know the values of several parameters. 
In this section, we demonstrate numerically obtained values
using the two-dimensional Lennard--Jones fluids at the hovering states of $\chi=12.65$ and $\Delta^{\rm L}\simeq 0.5 L_x$.
For a wide range of $N$, the number densities are obtained as $\rho^{\rm L}\simeq 0.74/\sigma^2$, $\rho_1^{\rm G}\simeq 0.03/\sigma^2$,
and $\rho_2^{\rm G}\simeq 1.1 \rho_1^{\rm G}$. 

For estimating $d P_{\rm s}/dT$, we start with the Clausius-Clapeyron relation 
\begin{align}
T\frac{dP_{\rm s}}{dT}=\frac{\hat q_{\rm s}}{\hat v_{\rm s}^{\rm G}-\hat v_{\rm s}^{\rm L}},
\label{e:Cla-Cla}
\end{align}
where $\hat q_{\rm s}$ is a latent heat and $\hat v_{\rm s}^{\rm G}$ and $\hat v_{\rm s}^{\rm L}$ are specific volumes of the gas and liquid at saturation.
The latent heat is the difference of the specific enthalpy between the saturated liquid and gas.
It is written as
$\hat q_{\rm s}=\hat u_{\rm s}^{\rm G}-\hat u_{\rm s}^{\rm L}+P_{\rm s}(\hat v_{\rm s}^{\rm G}-\hat v_{\rm s}^{\rm L})$
with specific internal energies $\hat u_{\rm s}^{\rm G}$ and $\hat u_{\rm s}^{\rm L}$ for the saturated liquid and gas.
For $\rho^{\rm L}\gg\rho^{\rm G}$, \eqref{e:Cla-Cla} is approximated as
\begin{align}
T\frac{dP_{\rm s}}{dT}\simeq \rho^{\rm G}(\hat u_{\rm s}^{\rm G}-\hat u_{\rm s}^{\rm L})+P_{\rm s}.
\label{e:dPs/dT}
\end{align}
As shown in Fig. 4, the values of the pressure are different between the hot and cold gases but the deviation is at most $10 \%$. We thus adopt the estimated value as  $P_{\rm s}\simeq 0.010\vep/\sigma^2$ corresponding to the mean pressure in the hot and cold gases.
Similar deviation is also observed in the specific internal energy $\hat u(x)=u(x)/\rho(x)$ shown in Fig. \ref{f:u-profile},
where $\hat u_{\rm s}^{\rm G}\simeq 0.1\vep$ and $-0.1\vep$ for the hot and cold gases, respectively.
We then adopt the mean value $\hat u_{\rm s}^{\rm G}\simeq 0.0\vep$ as the specific internal energy for the gas.
For the liquid, the spatial mean provides $\hat u_{\rm s}^{\rm L}\simeq -1.9\vep$.
Substituting these estimated values into \eqref{e:dPs/dT} yields 
\begin{align}
T\frac{dP_{\rm s}}{dT}\simeq 0.071 \vep/\sigma^2, 
\end{align}
where $\kB \Tm=0.43\vep$.

Figures \ref{f:conductivity} show the temperature profiles.
There are three regions when the liquid hovers.
The slope of temperature in each region is determined
as $J/\kappa$ with the heat flux $J$ and heat conductivity $\kappa$.
Since $J$ takes the same value for three layers, we obtain the
ratio of the heat conductivities as
\begin{align}
\frac{\kappa^{\rm L}}{\kappa_1^{\rm G}}=6.3,\quad
\frac{\kappa^{\rm L}}{\kappa_2^{\rm G}}=2.5
\end{align}
for $L_x/\sigma=1581$ with $N=5.0\times 10^5$,
and
\begin{align}
\frac{\kappa^{\rm L }}{\kappa_1^{\rm G}}=6.7,\quad
\frac{\kappa^{\rm L }}{\kappa_2^{\rm G}}=2.4
\label{e:ratio-kappa-80000}
\end{align}
for $L_x/\sigma=632$ with $N=8.0\times 10^4$.
Since the ratios of the heat conductivities
are comparable in the two system sizes,
we adopt the values in \eqref{e:ratio-kappa-80000} for the estimates.
Substituting the estimated parameter values into (9) in the main text, we obtain
\begin{align}
\chi_{\rm min}=7.8,  \quad 
\chi_{\rm max}=16.
\end{align}

Here, we note that heat conductivities in the hot and cold gasses
are different as expressed by $\kappa_2^{\rm G}/\kappa_1^{\rm G}\simeq 2.5$.
The difference is not explained by the small differences of the
temperature or the number density.
As an observation, the cold gas often contains tiny drops of the liquid.
In Fig. \ref{f:droplet}, one can see Brownian motion of liquid droplets inside the cold gas layer.
The droplets seem to survive until they are absorbed into the liquid.
In contrast, the droplets formed inside the hot gas layer evaporate rather
soon before showing the Brownian motion.
The Brownian motion of the liquid droplets could be typically observed in supercooled gases,
and it may modify the heat conductivity of the gas.

\subsection{Thermodynamic instability of the gas above the liquid}
\label{s:instability}

We deal with the cold gas situated in $x_2^{\rm int}<x<L_x$.
In order for the cold gas to be thermodynamically stable, the inequality
\begin{align}
P(x)\le P_{\rm s}(T(x))
\label{e:stableG}
\end{align}
should be satisfied.
Using the balance of force inside the cold gas 
\begin{align}
P_{\rm s}(T(x_2^{\rm int}))-P(x)=mg\rho^{\rm G}(x-x_2^{\rm int})
\end{align}
with the number density $\rho^{\rm G}$ for the gas,
the stability condition \eqref{e:stableG} leads to
\begin{align}
  -\frac{dP_{\rm s}}{dT}\frac{T(x)-T(x_2^{\rm int})}{x-x_2^{\rm int}}
  \le mg\rho^{\rm G},
\end{align}
where $dP_{\rm s}/dT$ is evaluated at $T=\Tm$ by repeating the
same argument just above \eqref{e:chi-Gamma-main}. 
Letting the heat conductivity for the cold gas be $\kappa_2^{\rm G}$, the steady heat flux is
\begin{align}
J=-\kappa_2^{\rm G}\frac{dT}{dx}.
\end{align}
We then have
\begin{align}
J\frac{dP_{\rm s}}{dT}\le mg\rho^{\rm G}{\kappa_2^{\rm G}}.
\label{e:rho-kappa-uG}
\end{align}
The relation \eqref{e:chi-Gamma-x} for the hovering liquid is transformed into
\begin{align}
J\frac{dP_{\rm s}}{dT} =mg \rho^{\rm L}\kappa^{\rm L}
\label{e:rho-kappa-L}
\end{align}
by applying the Fourier law.
Substituting \eqref{e:rho-kappa-L} into \eqref{e:rho-kappa-uG},
we conclude that the stability condition \eqref{e:stableG} for the cold gas is expressed as
\begin{align}
\rho^{\rm G}{\kappa_2^{\rm G}}\ge \rho^{\rm L}\kappa^{\rm L}.
\label{e:stability-cold-gas}
\end{align}
The inequality does not hold in the present numerical simulations of the Lennard--Jones fluids.
As mentioned in Sec. \ref{s:numerical estimates}, the numerical estimates give
\begin{align}
\frac{\rho^{\rm G}}{\rho^{\rm L}}\simeq 0.041, \quad
\frac{\kappa^{\rm L }}{\kappa_2^{\rm G}}\simeq 2.4,
\end{align}
which is out of the inequality \eqref{e:stability-cold-gas}. 
This is consistent with the numerical observation that the cold gas is supercooled.
We emphasize that the argument above follows even though the liquid does not float up and remains on the bottom of the container. 

The stability condition for the hot gas in $0<x<x_1^{\rm int}$ is formulated by  a parallel argument and expressed by
\begin{align}
\rho^{\rm G}{\kappa_1^{\rm G}}\le \rho^{\rm L}{\kappa^{\rm L}}.
\label{e:stability-hot-gas}
\end{align}
The numerical values satisfy this inequality, and therefore, the hot gas below the liquid is thermodynamically stable.

Note that the inequality \eqref{e:stability-cold-gas} is hard to hold in general far from the critical point because $\rho^{\rm L}\gg \rho^{\rm G}$.
For instance, the saturating water at atmospheric pressure would not satisfy the inequality.
Even though the heat conductivity is significantly larger in liquid than in gas, the number density is more different.
Thus, we expect that the gas on the saturating liquid is metastable in general.

\subsection{Local states in equilibrium phase diagram}
\label{s:local-meta}

\begin{figure}[bt]
\centering
\includegraphics[width=0.8\textwidth]{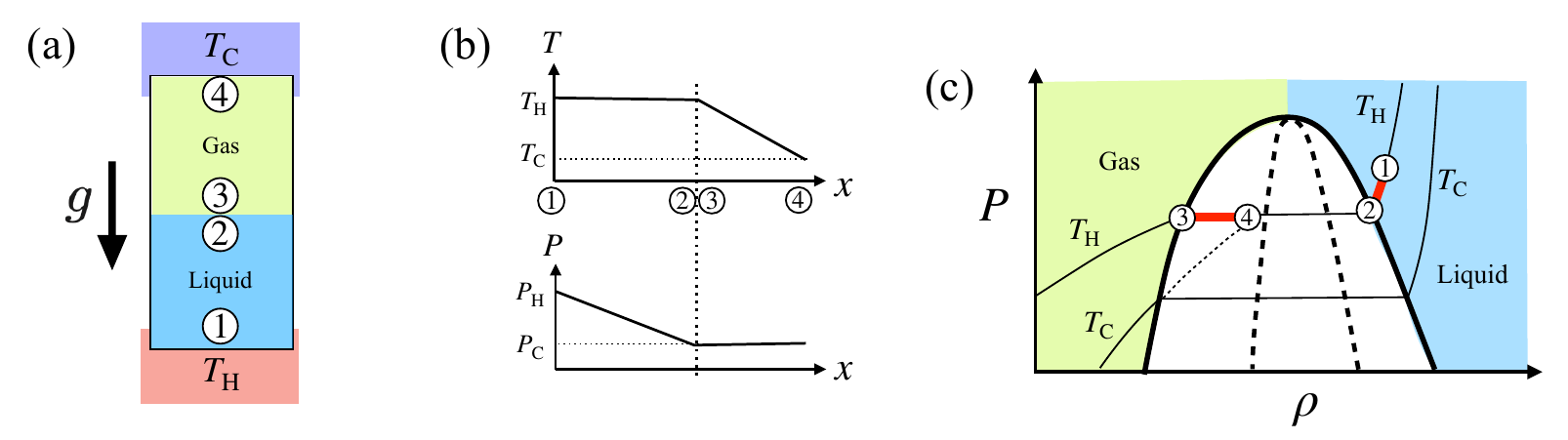}
\caption{
Supercooled gas induced by the heat flow under gravity for the limiting case $\kappa^{L} \gg \kappa^{G}$ and $\rho^{L} \gg \rho^{G} \simeq 0$.  See the text for the explanation.
(a) Configuration of liquid-gas coexistence when gas is placed above liquid. \textcircled{\scriptsize 1}, \textcircled{\scriptsize 2}, \textcircled{\scriptsize 3}, and \textcircled{\scriptsize 4} indicate 
the bottom, the liquid and gas sides of the interface, and the top.
(b) Profiles of temperature $T(x)$ and pressure $P(x)$ of the configuration in (a). The four marked points are indicated along the $x$-axis.
(c) All local states of the configuration in (a) are depicted in the equilibrium phase diagram as the red lines. 
Local states at \textcircled{\scriptsize 1}, \textcircled{\scriptsize 2}, \textcircled{\scriptsize 3}, and \textcircled{\scriptsize 4} in (a) are indicated.
Bold solid and bold dotted lines are the coexistence curve and the spinodal line. 
Two solid lines are the equations of state at $\HT$ and $\LT$ and a thin dotted line is metastable states at $\LT$.
The blue and green areas indicate thermodynamically stable states for liquid and gas.
}
\label{f:eos_meta}
\end{figure}

When the system is subjected to both gravity and heat flow, there are two constraints in the steady states:
the continuity of the pressure $P(x)$ and the uniformity of the heat flow in $x$.
Below, we demonstrate how these constraints produce the local states out of local equilibrium.
Let us consider the limiting case $\kappa^{\subL} \gg \kappa^{\subG}$ and $\rho^{\subL} \gg \rho^{\subG}\simeq 0$, 
where $\kappa^{\rm L/G}$ and $\rho^{\rm L/G}$ are the heat conductivity and the number density of the liquid or gas.

Suppose that the liquid and gas are distributed as shown in Fig.~\ref{f:eos_meta}(a), which is the stable configuration 
in equilibrium at $\HT=\LT$ under gravity.
\textcircled{\scriptsize 1}, \textcircled{\scriptsize 2}, \textcircled{\scriptsize 3}, and \textcircled{\scriptsize 4} indicate 
the bottom, the liquid and gas sides of the interface, and the top.
When $\HT=\LT$, the pressure in the gas layer is uniform at the saturation pressure $P_{\rm s}(\HT=\LT)$.
When $\HT>\LT$, the profiles $T(x)$ and $P(x)$ become as shown in Fig.~\ref{f:eos_meta}(b).
Because $\kappa^{\subL} \gg \kappa^{\subG}$, the temperature $T(x)$ is uniform at $\HT$ between \textcircled{\scriptsize 1} and \textcircled{\scriptsize 2} but sloped between \textcircled{\scriptsize 3} and \textcircled{\scriptsize 4}.
On the contrary, the pressure $P(x)$ is uniform between  \textcircled{\scriptsize 3} and \textcircled{\scriptsize 4}, say at $P_{\rm C}$,  but sloped in the liquid between \textcircled{\scriptsize 1} and \textcircled{\scriptsize 2} because  $\rho^{\subL} \gg \rho^{\subG}\simeq 0$.
Thus, the local state at the liquid-gas interface at \textcircled{\scriptsize 2} and  \textcircled{\scriptsize 3} is identified as $(\HT, P_{\rm C})$.

  We then assume $P_{\rm C}=P_{\rm s}(\HT)$, i.e., 
the pressure is saturated at \textcircled{\scriptsize 2} and  \textcircled{\scriptsize 3}.
The change of local states along $x$ is illustrated in Fig.~\ref{f:eos_meta}(c)
on the equilibrium phase diagram in $\rho$-$P$ space with the liquid-gas coexistence curve.
The two red lines represent the plots of all local states $(\rho(x),P(x))$ determined from the profiles in Fig.~\ref{f:eos_meta}(b).
The line connecting \textcircled{\scriptsize 1} and \textcircled{\scriptsize 2} represents the equation of state at $\HT$.
The points \textcircled{\scriptsize 2} and \textcircled{\scriptsize 3} are separated from each other, however,  they are both on the coexistence curve, one is on the liquid side and the other is on the gas side.
Note that the three points \textcircled{\scriptsize 2}, \textcircled{\scriptsize 3}, and \textcircled{\scriptsize 4} are on the line $P=P_{\rm C}$.
The temperature should decrease from $\HT$ at \textcircled{\scriptsize 3} to $\LT$ at \textcircled{\scriptsize 4} due to the Fourier law along $x$-axis.
Then, there is no choice for the local states other than being inside the coexistence curve as shown by the red line connecting \textcircled{\scriptsize 3} and \textcircled{\scriptsize 4}, that is, becoming supercooled gas.

\begin{figure}[bt]
\centering
\includegraphics[width=0.6\textwidth]{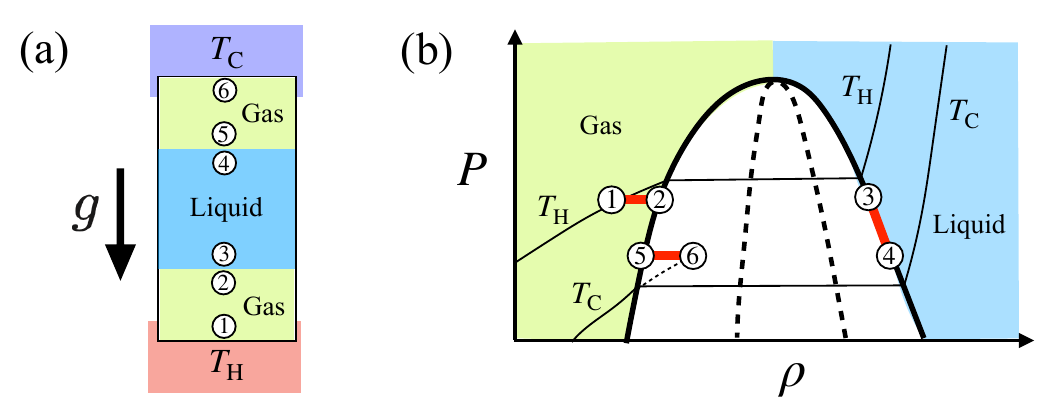}
\caption{
Local states in the hovering state described in Fig. 4 in the main text.
(a) Schematic figure for the configuration of the hovering state in Fig. 4.
 \textcircled{\scriptsize 1} indicates the bottom, 
\textcircled{\scriptsize 2} and \textcircled{\scriptsize 3} indicate the two sides of
the interface between the liquid and the hot gas,
while \textcircled{\scriptsize 4} and \textcircled{\scriptsize 5} do
the two sides of the interface between the liquid and the cold gas.
\textcircled{\scriptsize 6} indicates the top.
(b) All local states in Fig. 4 (red line) on the equilibrium phase diagram. Bold solid and bold dotted lines are the coexistence curve and the spinodal line. 
Two solid lines are the equations of state at $\HT$ and $\LT$ and a thin dotted line is metastable states at $\LT$.
The blue and green areas indicate thermodynamically stable states for liquid and gas.
}
\label{f:hovering_meta}
\end{figure}

Let us examine the hovering state with two interfaces, whose 
  local quantities are determined in Fig. 4 in the main text. 
The corresponding configuration is schematically shown in Fig.~\ref{f:hovering_meta}(a) 
with marked points at the bottom as \textcircled{\scriptsize 1}, 
at the two sides of the interface between the liquid and the hot gas as \textcircled{\scriptsize 2} and \textcircled{\scriptsize 3},
at the two sides of the interface between the liquid and the cold gas as \textcircled{\scriptsize 4} and \textcircled{\scriptsize 5},
and at the top as \textcircled{\scriptsize 6}. 
The local states are exhibited in Fig.~\ref{f:hovering_meta}(b)
on the equilibrium phase diagram in $\rho$-$P$ space with the liquid-gas coexistence curve.
According to the arguments in Sec. \ref{press-liquid}, 
   the local states in the liquid must be entirely on the coexistence curve as shown by the red line connecting \textcircled{\scriptsize 3} and  \textcircled{\scriptsize 4}.
The local states in the hot gas are on the red line connecting \textcircled{\scriptsize 1} with \textcircled{\scriptsize 2} inside the green area.
The local states in the cold gas are on the red line connecting \textcircled{\scriptsize 5} with \textcircled{\scriptsize 6} inside the coexistence curve.
Thus, the local states in the cold gas are not thermodynamically stable, whereas those in the hot gas are thermodynamically stable. The cold gas is supercooled.

\section{Setups of experiments for noble gases}
\label{s:experimental}

As demonstrated in Sec. \ref{s:hovering}, the liquid floats up regardless of the details of the configuration in two-dimensional systems
and therefore, we assume that this is also the case in three-dimensional systems.
Because noble gases are modeled by the mono-disperse Lennard--Jones systems,
the numerical scaling shown in Fig. 3 of the main text can provide the estimate for the setups to observe the phenomena in real experiments.

The definition of $\chi$ in (6) of the main text leads to the mean temperature gradient for the system as 
\begin{align}
\frac{\Delta T}{L_x}=\frac{mg\chi}{\kB}
=\frac{m_* g \chi}{8.31\times  10^3} ~~\mathrm{[K/m]},
\label{e:estimate1}
\end{align}
where $m_*$ is the molar mass, $m_*=m N_{\rm A} \times 10^3$ [g/mol] with the Avogadro number 
$N_{\rm A}=6.02\times 10^{23}$ and $\kB=1.38\times 10^{-23}  \, \mathrm{J/K}$.

Figure 1 in the main text corresponds to the three-dimensional example 
with $\chi=8.37$, $\rho\sigma^3=0.30$, and $\kB \Tm/\vep=1.0$.
We now suppose that $\chi=100$ is a typical value for observing the hovering states of real noble gases
when the number density and the temperature correspond to  $\rho\sigma^3=0.30$ and $\kB \Tm/\vep=1.0$.
Substituting 
$g=9.8 \, \mathrm{m/s^2}$ and $\chi=100$
into \eqref{e:estimate1} yields
\begin{align}
\frac{\Delta T}{L_x}=0.12~ m_*\quad \mathrm{[K/m]},
\end{align}
and 
\begin{align}
\LT=\Tm-0.060~ m_* L_x ~~\mathrm{[K]}, \qquad \HT=\Tm+0.060~ m_* L_x ~~\mathrm{[K]}.
\label{e:TL-TH}
\end{align}
The value of $\Tm$ should be set around the liquid-gas transition temperature.

For instance, xenon is simulated by the Lennard--Jones system with $m=2.180\times 10^{-25}$ kg, $\vep/\kB=2.182\times 10^2$ K, and $\sigma=4.055\times 10^{-10}$ m.
These parameter values provide 
$\rho=7.5 ~ \mathrm{mol/l}$ for $\rho\sigma^3=0.30$
and $\Tm=218.20$ K for $\kB \Tm=\vep$.
With the molar mass $m_*=131$ and $L_x=0.01$ m, we have $\LT=218.12$ K and $\HT=218.28$ K.
We do similar calculations to other species from the Lennard--Jones parameters and obtain Table  \ref{t:noblegas} for the experimental setups.

\begin{table}[h]
\centering
\begin{tabular}{| c | c | c | c | c |}
\hline
&~~~~ Ne ~~~~ &~~~~  Ar ~~~~ & ~~~~ Kr ~~~~ & ~~~~ Xe~~~~ \\
\hline
$m_* ~~\mathrm{[g/mol]}$ & 20 & 40 & 84&131\\
\hline
$\mT~~\mathrm{[K]}$ & 36.83 &116.8&164.6&218.2\\
\hline
~~$m\rho ~~\mathrm{[kg/l]}$~~  &0.47 &0.51& 0.89 & 0.98 \\
\hline\hline
$\sigma ~~\mathrm{[\AA]}$ & 2.775& 3.401 &3.601 & 4.055\\
\hline
\end{tabular}
\caption{Molar mass $m_*$, temperature $\mT$ chosen as the liquid-gas coexistence, and  mass density $m\rho$
for each noble gas for utilizing \eqref{e:TL-TH}. These values are determined from the parameters of Lennard--Jones systems with $\mT=1.0\vep/\kB$ and $\rho=0.30 \sigma^3$. The values of $\vep$ and $\sigma$ are taken from \cite{oh2013_}.}
\label{t:noblegas}
\end{table}

\end{document}